\documentclass[a4paper,11pt]{article}
\pdfoutput=1 

\usepackage{jcappub} 

\usepackage[T1]{fontenc} 

\usepackage{graphicx}
\usepackage{amssymb}
\usepackage{lineno}
\usepackage{units}
\usepackage[super]{nth}
\usepackage{color}
\usepackage{siunitx}

\title{A search for cosmogenic neutrinos with the ARIANNA test bed using 4.5 years of data}

\author[a]{A.~Anker}
\author[a]{S.~W. Barwick}
\author[b]{H. Bernhoff}
\author[c,d]{D.~Z. Besson}
\author[e]{N. Bingefors}
\author[f,g]{D. Garc\'ia-Fern\'andez}
\author[a]{G. Gaswint}
\author[a]{C. Glaser}
\author[e]{A. Hallgren}
\author[h]{J.~C. Hanson}
\author[i]{S.~R. Klein}
\author[j]{S.~A. Kleinfelder}
\author[a,g]{R. Lahmann}
\author[c]{U. Latif}
\author[k]{J. Nam}
\author[c,d]{A. Novikov}
\author[f,g]{A. Nelles}
\author[a]{M.~P. Paul}
\author[a,1]{C. Persichilli \note{Corresponding author}}
\author[f,g]{I. Plaisier}
\author[i]{T. Prakash}
\author[a]{S. R. Shively}
\author[a,l]{J. Tatar}
\author[e]{E. Unger}
\author[k]{S.-H. Wang}
\author[f, g]{C. Welling}

\affiliation[a]{Department of Physics and Astronomy, University of California, Irvine, CA 92697, USA}
\affiliation[b]{Uppsala University Department of Engineering Sciences, Division of Electricity, Uppsala, SE-752 37 Sweden}
\affiliation[c]{Department of Physics and Astronomy, University of Kansas, Lawrence, KS 66045, USA}
\affiliation[d]{National Research Nuclear University MEPhI (Moscow Engineering Physics Institute), Moscow 115409, Russia}
\affiliation[e]{Uppsala University Department of Physics and Astronomy, Uppsala, SE-752
37, Sweden}
\affiliation[f]{DESY, 15738 Zeuthen, Germany}
\affiliation[g]{ECAP, Friedrich-Alexander-Universit\"at Erlangen-N\"urnberg, 91058 Erlangen, Germany}
\affiliation[h]{Whittier College Department of Physics, Whittier, CA 90602, USA}
\affiliation[i]{Lawrence Berkeley National Laboratory, Berkeley, CA 94720, USA}
\affiliation[j]{Department of Electrical Engineering and Computer Science, University of California, Irvine, CA 92697, USA}
\affiliation[k]{Department of Physics and Leung Center for Cosmology and Particle Astrophysics, National Taiwan University, Taipei 10617, Taiwan}
\affiliation[l]{Research Cyberinfrastructure Center, University of California, Irvine, CA 92697 USA}

\emailAdd{cpersich@uci.edu, sbarwick@uci.edu, christian.glaser@uci.edu}

\abstract{
The primary mission of the ARIANNA ultra-high energy neutrino telescope is to uncover astrophysical sources of neutrinos with energies greater than \SI{e16}{eV}.  A pilot array, consisting of seven ARIANNA stations located on the surface of the Ross Ice Shelf in Antarctica, was commissioned in November 2014. We report on the search for astrophysical neutrinos using data collected between November 2014 and February 2019.  A straight-forward template matching analysis yielded no neutrino candidates, with a signal efficiency of $79\%$. We find a 90\% confidence upper limit on the diffuse neutrino flux of $E^2\Phi=\SI{1.7e-6}{GeV cm^{-2}s^{-1}sr^{-1}}$ for a decade wide logarithmic bin centered at a neutrino energy of \SI{e18}{eV}, which is an order of magnitude improvement compared to the previous limit reported by the ARIANNA collaboration. The ARIANNA stations, including purpose built cosmic-ray stations at the Moore's Bay site and demonstrator stations at the South Pole, have operated reliably. Sustained operation at two distinct sites confirms that the flexible and adaptable architecture can be deployed in any deep ice, radio quiet environment. We show that the scientific capabilities, technical innovations, and logistical requirements of ARIANNA are sufficiently well understood to serve as the basis for large area radio-based neutrino telescope with a wide field-of-view.
}

\begin{document}
\maketitle
\flushbottom





\section{Introduction}
\label{sec:Introduction}

Since the discovery of extremely energetic cosmic rays more than a half century ago, the elusive quest to uncover the sources of these enigmatic particles has provided many challenges. Despite progress in experimental capabilities and theoretical insight, we do not yet know the acceleration mechanism for those particles with energies that have been measured in excess of \SI{e20}{eV} \cite{AugerSpectrumICRC2017}. Being electrically charged, cosmic rays are deflected by galactic and intergalactic magnetic fields such that the arrival direction at the Earth no longer points back to their origin, making source identification difficult. In addition, interactions with cosmic microwave photons limit the direct observation of ultra-high energy cosmic rays sources to our local supercluster (the GZK horizon) \cite{Greisen1966,Zatsepin1966}.  

Neutrino astronomy offers a new and powerful tool to provide insight into the physics associated with the acceleration process, and complements and extends measurements not accessible through the observation of other messengers. Charged cosmic rays which interact with gas, dust, or radiation near an accelerating object produce $\gamma$-rays and high-energy neutrinos; referred to here as astrophysical neutrinos. Whereas $\gamma$-rays can be absorbed in dense environments, these astrophysical neutrinos can escape and travel unimpeded to a detector (\cite{DecadalWhitePaper} and references therein). Neutrinos effectively propagate at the speed of light and, unlike charged cosmic rays, neutrinos are not deflected by magnetic fields; which allows for identification of sources, as well as directional and temporal coincidence with photons and gravitational waves from the source object.

The most energetic cosmic rays which do escape their source can interact with the cosmic microwave background en route to the Earth,  generating cosmogenic neutrinos with a characteristic energy distribution peaking at \SI{e18}{eV} \cite{Stecker1973,Beresinsky1969}. Since the interaction lengths for these neutrinos through the cosmic microwave background would be larger than the observable universe, the detection of neutrinos originating well outside the GZK horizon of \SI{100}{Mpc} becomes possible.  

Since neutrinos interact very weakly with matter, and the expected fluxes at the relevant energies are low, large volumes of detection medium are necessary to make a significant measurement. The glaciers of Antarctica and Greenland provide a natural target, and have been leveraged in several detectors including IceCube, ANITA, and RICE \cite{IceCubeLimit2018,Gorham2018,RICELimit2012}.  The  best limit on the ultra-high energy (UHE, $E_\nu > \SI{e16}{eV}$) neutrino flux up to approximately $\SI{e20}{eV}$ is currently set by the IceCube collaboration \cite{IceCubeLimit2018}.  However, this limit is still orders of magnitude away from contesting the most conservative models of the cosmogenic flux.  

The evidence for astrophysical neutrinos in the energy interval between \SI{10}{TeV} and \SI{10}{PeV} has grown strong over the last decade \cite{kopper2017observation,IcecubeMuonFlux2017,IceCubeLimit2018}. The arrival directions do not cluster around galactic matter but rather follow a uniform distribution indicating an extragalactic origin.
Furthermore, the intriguing spatial coincidence of high energy neutrinos with the blazar TXS 0506+056 \cite{IceCubeBlazar2018,IceCubeBlazarCorrelation2018} reported by the IceCube Collaboration prompted renewed interest in the role of point sources in multi-messenger astronomy.  Theories of UHE cosmic-ray production predict a deep relationship between gamma-rays, astrophysical neutrinos, and UHE charged cosmic rays  \cite{Ahlers2018}.  The ability of future radio-based neutrino detectors to measure luminous transient or variable events will be a key requirement for multi-messenger astronomy, and the large effective volumes made possible by radio neutrino detectors provide a low cost opportunity to discover rare or unexpected explosive sources.

While the current paucity of neutrino point sources for neutrino energies between TeV to PeV suggests a population of weak extragalactic sources, it remains an open question if this conclusion is valid at much greater neutrino energies.  Perhaps rarer, more luminous, sources will be the keys to understanding the generation of the highest energy cosmic rays.  There is growing recognition that "the spatial distribution and clustering of high-energy neutrinos across the sky are key observables for revealing their origins" \cite{DecadalWhitePaper}. This requirement prioritizes sky coverage and pointing resolution for every neutrino event. The ARIANNA \cite{ARIANNAPrototype,2015ARIANNALimitsPaper,COSPAR2019} concept achieves large sensitivity by minimizing logistical and component costs, simplifying system complexity, and reducing data management costs. Equally important, the ARIANNA design broadens the science capabilities through accurate direction and energy reconstruction, which are discussed in detail in \cite{GlaserDNR} and \cite{GlaserICRC2019}.

Though experiments such as IceCube will continue to improve the limits on diffuse flux by continuing to accrue livetime, sensitivity to transient point sources only depends on effective volume and backgrounds, which create interesting opportunities for radio-based neutrino observatories.  The virtues of the ARIANNA concept have been discussed previously \cite{2015ARIANNALimitsPaper}, and a summary of the detector design can be found in Sec.~\ref{sec:Detector}. By taking advantage of $\mathcal{O}(\SI{1}{km})$ attenuation lengths of radio signals in the relevant frequency range in glacial ice, a sparse grid of radio detectors can feasibly instrument hundreds of \SI{}{km^3}, which is difficult for an optical (IceCube style) detector \cite{CollaborationAartsenAbrahamEtAl2016}.

There are several pilot-stage projects which are exploring contrasting designs for the radio-based detection of ultra-high energy (UHE) neutrinos.  ARA \cite{ARALimit2016}, for example, is an in-situ radio array located near the South Pole.  The ARA design employs antennas which are placed in boreholes $\sim\SI{200}{m}$ below the snow surface.  Locating the antennas below the firn allows for a slightly greater field of view than a surface detector at the same location (site location impacts field of view, as discussed in Sec.~\ref{sec:SiteSelect}).

Other proposed designs would aspire to measure a single flavor flux of tau neutrinos.  A $\nu_\tau$ which undergoes a charged current interaction within the crust, or a mountain, produces a $\tau$ lepton which may decay within the atmosphere. The decaying high-energy $\tau$ would then produce a measurable radio signal through the same mechanisms as a typical cosmic-ray air shower. Projects which aim to leverage this effect include GRAND \cite{GrandArXiv}, TAROGE \cite{JiwooICRC2019} and BEACON \cite{WisselICRC2019}. The ARIANNA site at Moore's Bay is also suited for this detection channel, as it is adjacent to  a large mountain range. A prototype tau neutrino station has been running successfully since 2017 \cite{Shih-HaoICRC2019}.

It may be non-obvious that a neutral particle should produce a strong radio signal, but the interaction of UHE neutrinos in ice produce particle showers containing millions of charged particles. Such showers produce coherent radio emission through the Askaryan effect \cite{Askaryan1965}, which has been experimentally verified in ice \cite{ANITA2007}. Interactions between the shower and the dense medium produce a time-varying charge excess along the shower front and near the shower axis.  This macroscopic charge excess leads to coherent radio emission concentrated within a few degrees of the Cherenkov cone angle, which is $55.8^\circ$ in deep ice with an index-of-refraction of $n=1.78$. This signal takes the form of an $\approx \SI{1}{ns}$ bipolar pulse with broad-band frequency content exceeding $\SI{1}{GHz}$.

\section{The ARIANNA Detector}
\label{sec:Detector}

The ARIANNA test bed of radio-based neutrino stations is primarily located in Moore's Bay on the Ross Ice Shelf, with two additional demonstrator stations located at the South Pole. Each ARIANNA station is designed to operate autonomously, with self-contained power, data acquisition (DAq), and communication systems.  A single station is capable of identifying and reconstructing a neutrino signal without the need for multi-station coincidence. This allows stations to be deployed in any configuration at multiple sites, providing a straightforward path towards large area and full sky coverage. In this section, we will discuss the hardware, design philosophy, and capabilities of the detector. For a more thorough description, see \cite{ChrisThesis}.

\subsection{The 7 Station Test Bed}

The main goals of the now complete ARIANNA pilot program were to:
\begin{enumerate}
    \item Assess the durability and efficacy of the ARIANNA architecture.
    \item Evaluate the radio noise environment.
    \item Study the propagation of waves at radio frequencies through the polar ice medium.
    \item Gain experience with deployment and other logistical issues.
\end{enumerate}

To address these goals, a total of seven ARIANNA stations were deployed on the surface of Moore's Bay on the Ross Ice Shelf in Antarctica and successfully operated for more than four years.  The first four of the seven stations \cite{2015ARIANNATechnologyPaper} that make up the current test bed were deployed in the 2014-2015 Antarctic summer season, with the remaining three deployed the following season, replacing the previous generations of stations that was analyzed in \cite{2015ARIANNALimitsPaper}.

\subsection{The ARIANNA Site}
\label{sec:SiteSelect}

The ARIANNA site at Moore's Bay provides several unique advantages for the test bed. At a thickness of \SI[separate-uncertainty = true]{576(8)}{m} \cite{hanson2015}, the ice shelf at Moore's Bay provides sufficient volume for the interaction of UHE neutrinos. The low flow rate at Moore's Bay yields ice which is remarkably free of crevasses, and is expected to limit the effects of birefringence.  The depth averaged attenuation length, as a function of frequency, was measured to be $\langle L(f) \rangle = \SI[separate-uncertainty = true]{460(20)}{m} - \SI[separate-uncertainty = true]{180(40)}{m/GHz} \times f$ between the relevant frequency range of \SI{100}{MHz} to \SI{800}{MHz} assuming a reflection coefficient at the bottom of the ice shelf of $1$, making the measurement conservative (i.e., attributing all losses to ice attenuation) \cite{hanson2015}.

In addition, the ARIANNA stations are designed to take advantage of the boundary between the ice shelf and the Ross Sea beneath.  This boundary acts as a highly efficient mirror with an electric field reflection coefficient of $\sqrt{R} = 0.82 \pm 0.07$ \cite{hanson2015}, reflecting radio signals from neutrino interactions back up towards the surface mounted receivers. Correcting the attenuation length measurement for this reflection coefficient results in a depth-averaged attenuation length of approximately \SI{500}{m} at low frequencies.

\begin{figure}
    \centering
    \includegraphics[width=0.7\textwidth]{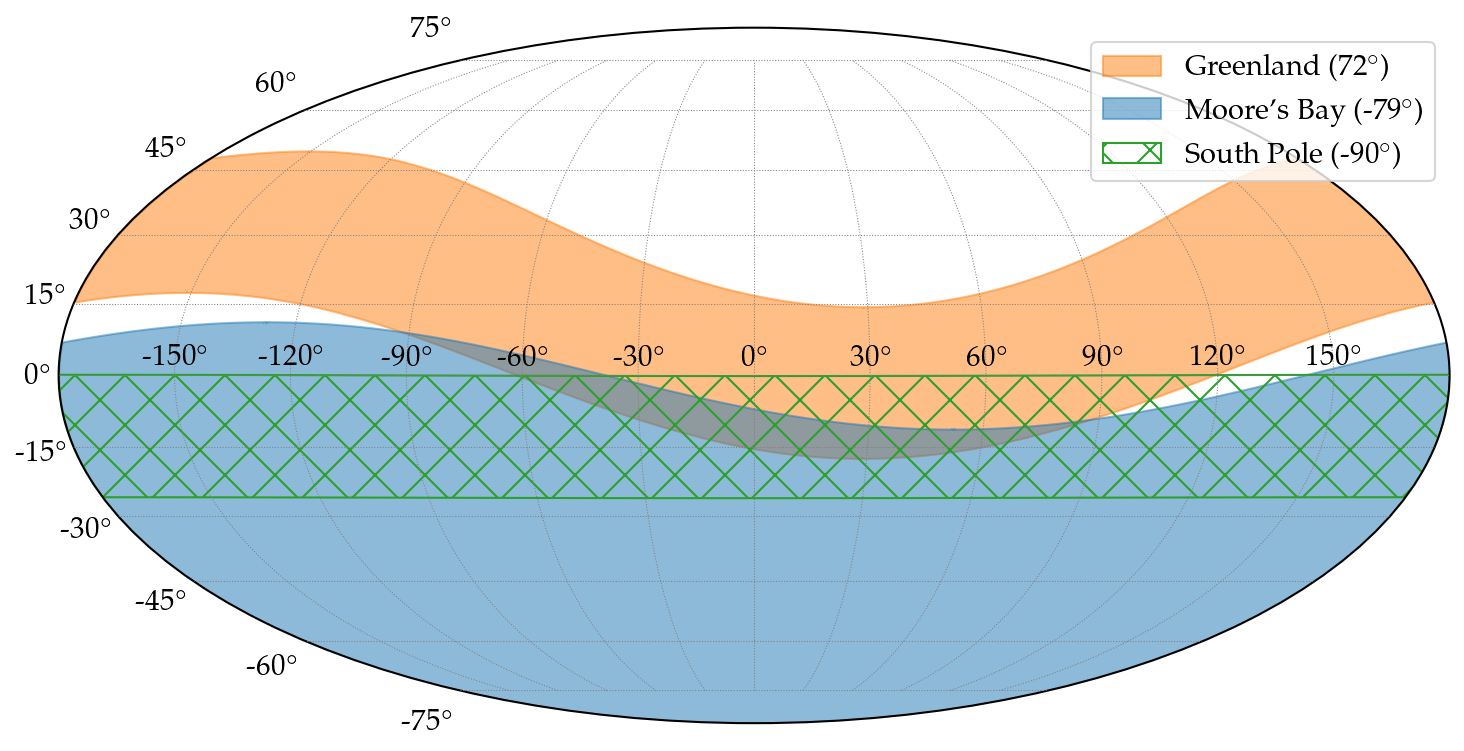}
    \caption{Instantaneous sky coverage for different radio detector locations in declination and right ascension.  The field of view is defined to cover the solid angle which contain 90\% of triggering events for an isotropic flux. The sky coverage was calculated for a \SI{60}{m} deep detector at the South Pole (green hash), for a \SI{50}{m} deep detector at Greenland (orange solid), and for an ARIANNA station at Moore's Bay (solid blue), including the reflections at the ice-water interface. Over \SI{24}{h}, the sky-coverage bands for Moore's Bay (at \SI{-79}{\degree} latitude) and Greenland (at \SI{72}{\degree} latitude) rotate horizontally due to the rotation of the Earth, increasing the field-of-view. In contrast, a detector at the South Pole always sees the same part of the sky.}
    \label{fig:SkyCoverage}
\end{figure}

Because of this unique feature, ARIANNA stations at Moore's Bay can view half the sky at any given instant in time with approximately uniform angular sky coverage (cf. Sec.~\ref{sec:MonteCarlo}). Due to the latitude of the ARIANNA site, different parts of the sky are exposed as the Earth rotates, increasing the field-of-view for sources which are sustained more than a few hours. The reflective layer is a unique feature in contrast to other locations, such as the South Pole and Greenland, where the field-of-view is limited to a $\sim\SI{30}{\degree}$ band in declination (see Fig.~\ref{fig:SkyCoverage}) at any given moment.

A future detector at Moore's Bay would complement IceCube in the search for point sources in the southern sky, and provide comparable or better angular resolution where events are dominated by shower topologies. Since ARIANNA's field-of-view extends approximately $11^\circ$ into positive declination, it overlaps a region of the sky which contains the highest energy tracks observed by IceCube.  For example, the declination band of ARIANNA includes the neutrino event observed by IceCube from the blazar TXS 0506+056 \cite{IceCubeBlazar2018}, a region of the sky unavailable to a radio detector at the South Pole. A detector at Moore's Bay would also scan over parts of the sky covered by optical neutrino telescopes under construction in the northern hemisphere \cite{KATZ_KM3Net,AVRORIN_GVD}.

\begin{figure}
  \begin{centering}
    \includegraphics[width=0.5\textwidth]{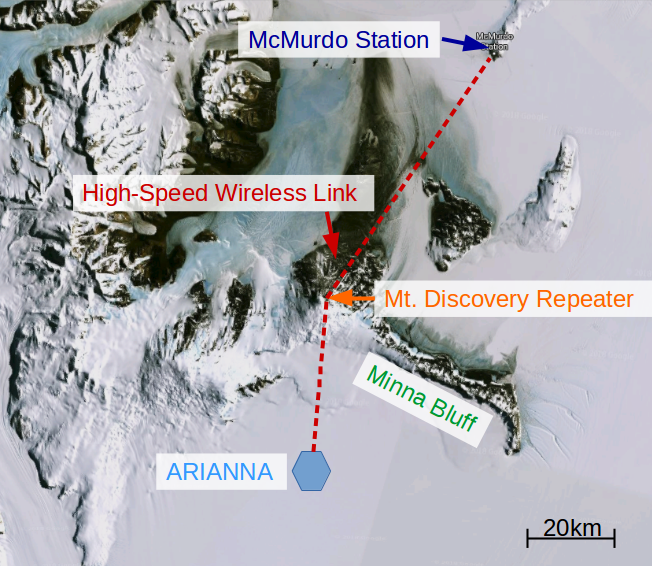}
    \caption{Location of the ARIANNA site in Moore's Bay (array size is not to scale).  Imagery courtesy of the U.S. Geological Survey, via Google Maps \cite{GoogleMaps}.}
    \label{fig:AriannaSiteSatellite}
  \end{centering}
\end{figure}

At the same time, the site is shielded from anthropogenic radio frequency (RF) noise originating from McMurdo operations by the nearby mountain ranges, especially Minna Bluff (see Fig.~\ref{fig:AriannaSiteSatellite}). This feature, along with its distance from McMurdo Station, makes the site remarkably radio quiet. 

At a distance of \SI{110}{km}, the ARIANNA site is relatively close proximity to the largest U.S. research base in Antarctica (McMurdo Station) allows for a variety of logistical support options, ranging from short-haul helicopter flights (used to transport personnel, cargo, tents, and fuel) to overland traverse (an option for a large scale future array).

\subsection{Communication and Data Transfer}
The geographical proximity to McMurdo Station enabled the installation of a bidirectional high-speed wireless link to McMurdo (Fig.~\ref{fig:AriannaSiteSatellite}), which maintains internet access 24/7.  Each station contains a long range AFAR pulsAR \SI{2.4}{GHz} wireless ethernet bridge \cite{AFAR} for near real-time data transfer.  When ARIANNA stations connect over this ethernet link, they will transfer all data from the previous data taking window, which is typically set to a 30 minute period. This system is capable of transmission speeds of up to \SI{200}{kB/s}, with each event taking approximately \SI{2}{kB}.

ARIANNA stations also contain an Iridium satellite modem for communication via the short burst data (SBD) protocol. This provides a redundant command and (limited) data transfer channel for the ARIANNA test bed. While the ethernet repeater  on Mt.\ Discovery (Fig.~\ref{fig:AriannaSiteSatellite}) operates on solar power and is only available during the summer, the Iridium network is available year-round.

ARIANNA stations operate asynchronously, periodically communicating with the control server at University of California-Irvine to transmit data and/or receive new configuration instructions. All data is saved on an SD card on the main DAq board, so data which is not transferred immediately can be later retrieved in a bulk data transfer, or physically recovered.

\subsection{Antennas}

\begin{figure}
  \centering
  \includegraphics[width=0.6\textwidth]{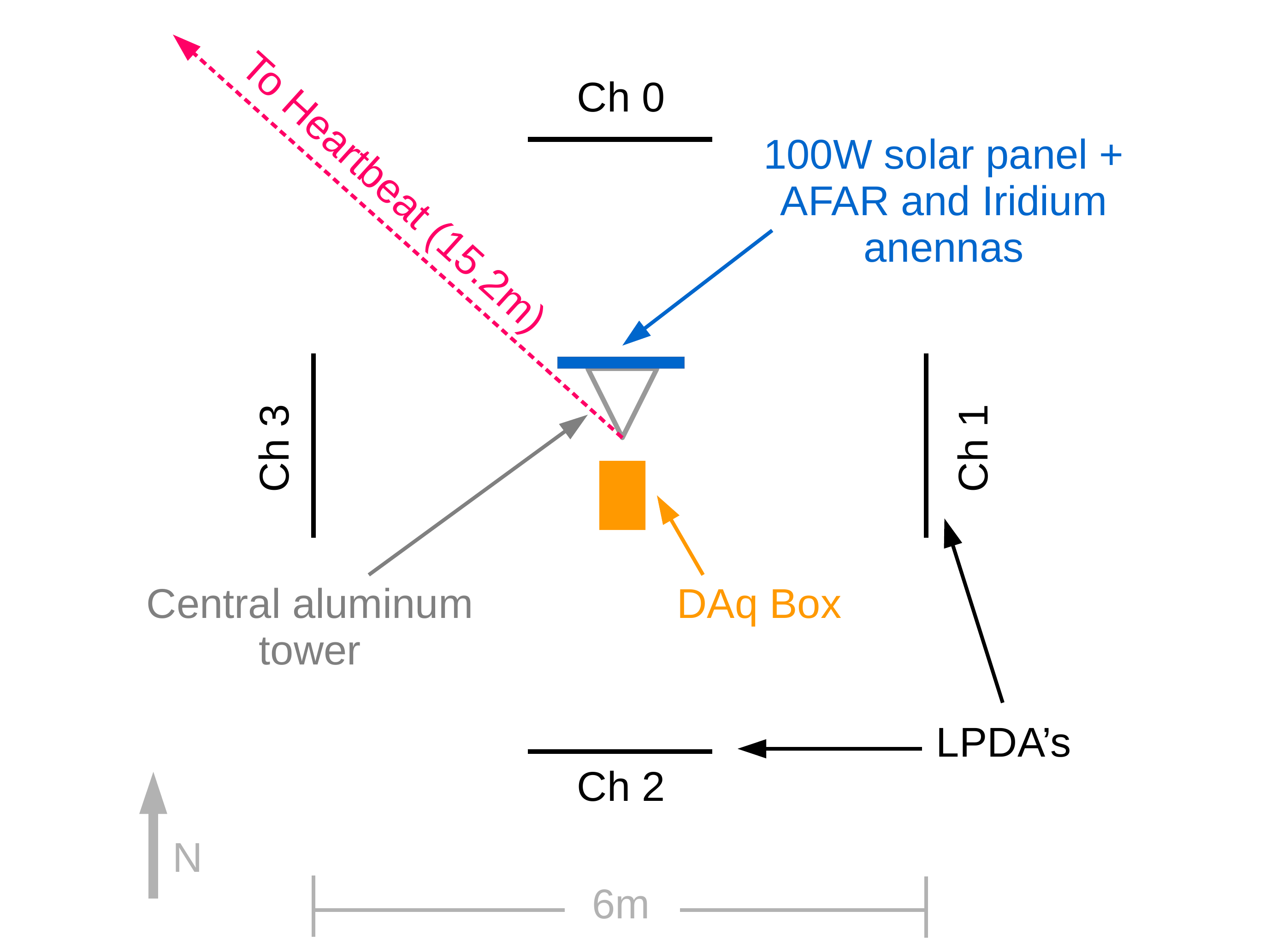}
  \caption{Top-down view of the ARIANNA station layout.  The southern vertex of the triangular tower acts as a reference for station center. Next to the 7 neutrino stations of this configuration, ARIANNA has five additional stations in various configurations, which are illustrated in \cite{COSPAR2019}. }
  \label{fig:StationTopDownView}
\end{figure}

The main detection element of each station consists of four Create Design Corp.\ CLP5130-2 log-periodic dipole antennas (LPDA) \cite{Create}.  These are arranged, buried face-down just below the snow level, in two pairs with perpendicular orientation, at a \SI{3}{m} radius from the station center (see Fig.~\ref{fig:StationTopDownView}).  These antennas have an effective bandwidth of 80 - \SI{1000}{MHz} when buried at the top of the firn \cite{ARIANNAPrototype}.  Each test bed station also contains a "heartbeat" LPDA placed \SI{15.2}{m} from the station center, which can direct a calibration pulse towards the detector.

\subsection{Data Acquisition}

The ARIANNA DAq mainboard is designed around the Synchronous Sampling plus Triggering (SST) chip  \cite{2015ARIANNATechnologyPaper, TarunThesis, EdwinThesis}.  Incoming signals are sampled at \SI{2}{GSa/s} and held in a 256 sample circular buffer, which is read out when a triggering condition is met. The SST has a remarkably accurate time synchronization of less than \SI{5}{ps} between samples and between channels.  This timing precision is invaluable for a precise event reconstruction, given the stations' modest \SI{6}{m} antenna separation (see \cite{COSPAR2019} and \cite{GaswintICRC2019} for an example event reconstruction using the ARIANNA DAq).

The SST employs a multi-stage coincidence trigger in order to dramatically reduce trigger rates from thermal noise.  The first stage requires a \SI{5}{ns} coincidence between high and low threshold crossing on a single input channel. Both the high and low threshold are adjustable separately per channel.  The second stage requires that two or more channels trigger within a \SI{30}{ns} window, taking advantage of the expectation that parallel pairs of LPDA's should see largely the same signal. A level one (L1) trigger can also be applied on triggered events after readout,  to filter out narrow-band anthropogenic background. The L1 trigger has a negligible effect on sensitivity, with a rejection rate of $1.8 \times 10^{-5}$ on simulated neutrino events \cite{PersichilliICRC2017}.

The ARIANNA electronics can save data to SD card at a maximum event rate of \SI{75}{Hz} for the 4 channel configurations (and half that rate for the 8 channel configurations), though more typically the trigger rate ranges between $10^{-3}$ to \SI{e-2}{Hz} corresponding to a trigger threshold of four times the RMS noise ($4 \times V_\mathrm{RMS}$). These rates were selected to reduce the time required to transmit data over the communication links.  Since the ARIANNA station cannot search for neutrino events during data transmission, this strategy reduced the deadtime of the detector. 

During most of the data taking time, the triggers are dominated by thermal noise fluctuations. Thus, the typical trigger rate is a function of trigger threshold, where the global trigger rate increases by roughly an order of magnitude when the trigger threshold is lowered by $0.13 \times V_\mathrm{RMS}$. Hence, the detector sensitivity depends only weakly on the trigger rate. Low trigger rates come with the advantage of real-time data transfer off-continent and allows modest technical requirements for data management and archiving.

In addition to the threshold trigger, ARIANNA stations generates a collection of minimum bias events by periodically forcing an event readout (typically once every \SI{67}{seconds}). For the study of background conditions as much as possible of this minimum bias data is transmitted in real-time.

\subsection{Amplifiers}
\label{sec:amps}

Each input channel contains a custom designed amplifier with a gain of roughly $\SI{60}{db}$ over a bandwidth of 100 - \SI{1000}{MHz}.  Each amplifier consists of four AC-coupled stages, each composed of a Avago MGA-68563 chip \cite{2015ARIANNATechnologyPaper}. There are two amplifier revisions deployed with the ARIANNA test bed, referred to as "100 series" and "200 series".  The 200 series was developed in order to reduce cost and weight, and was also adjusted for a flatter gain in the bandwidth of interest. As these two amplifier types have slightly different impulse response, they are treated separately in the analysis.  All versions of the ARIANNA amplifiers begin to clip at approximately \SI{800}{mV}, which limits the dynamic range of the system.

\subsection{Power Systems}
All ARIANNA stations in Moore's Bay are solar powered, with a cold weather optimized $LiFePO_4$ battery system.  The battery system can power a station for approximately 3 days without sun, and significantly increases the stations uptime during the shoulder seasons when the sun sets each day. The current battery system, along with a custom configured charge controller, was deployed on all stations during the 2015-2016 field season, and lead to a significant improvement in reliability and livetime \cite{PersichilliICRC2017}.

There has also been an effort to develop and operate wind power systems which can survive Antarctic conditions.  This would potentially allow remote ARIANNA stations to operate through winter without access to a power grid, greatly increasing the livetime of the array.  Prototype wind generators have been installed on one of the test bed stations, and have shown promising results.  The most recent prototype, deployed in November 2018, reached an operational livetime of 40\% for a station without solar panels. We estimate that it is feasible to
achieve at least 95\% livetime for a radio array running solely on renewable energies by scaling the turbine and battery size. See \cite{ICRC2019wind} for more details.

\subsection{Cosmic Ray Stations}

In addition to the 7 station test bed, several specialized stations were deployed at the Moore's Bay site to answer specific questions associated with the detection and identification (or tagging) of cosmic-ray-induced extended air showers \cite{NellesCRs2017}, which generate radio pulses of similar strength, frequency content, and duration as expected for signals from neutrinos. These stations comprise four (additional) upward facing LPDA's. The more abundant radio signals from air showers are not only an important background to tag, but also offer a unique way for an in-situ calibration and test of the detector. Using cosmic-ray data we already demonstrate a resolution in the polarization reconstruction of \SI{7}{\degree} \cite{AnnaICRC2019}, as well as measurement of the viewing angle, i.e., the angle between the shower axis and the line-of-sight to the emission point \cite{Welling2019, AnnaICRC2019}. These are crucial properties to determine the neutrino direction and energy, respectively \cite{GlaserICRC2019}.

Another station was deployed to measure cosmic rays arriving from directions close to the horizon \cite{Shih-HaoICRC2017,Shih-HaoICRC2019}. The goal of this configuration was to measure the flux at large zenith angles and evaluate the design as a first step toward a  tau neutrino detector. 

\subsection{South Pole Demonstrator Stations}
\label{sec:SouthPole}

The autonomous design of the ARIANNA system allows for a station to be deployed essentially anywhere there is ice of sufficient quality. To demonstrate this capability,  two ARIANNA detector stations were deployed near the South Pole Station. 

The first South Pole demonstrator, SP-1, was deployed in December of 2017. The station is located at a distance of approximately \SI{2.5}{km} from the main Amundsen-Scott Station. The SP-1 station receives DC power from the ARA \cite{Allison2012} electrical grid, allowing it to operate throughout the year. The power is routed approximately \SI{20}{m} from a junction box at the base of ARA Wind Turbine 3. 

In December 2018 a second station was deployed at a more remote location, approximately \SI{5}{km} from the main station.  This detector, SP-2, is fully independent, operating on a solar power system like the stations at Moore's Bay.  This station was designed to test the feasibility of operating a solar/battery power system in the colder temperatures at the South Pole, and performed reliably while sunlight was available.

\section{Data-set and Simulation}
\label{sec:data}

This section will define the data set to be used in this analysis (Sec.~\ref{sec:analysis}), as well as provide detail on the determination of the livetime used in the calculation of the limit of the diffuse neutrino flux (Sec.~\ref{sec:FluxLim}). We also discuss the Monte Carlo simulations that are used to calculate the detector effective volume, and construction of the simulated neutrino signal.

\subsection{Definition of the Data Set}

The analysis will concern the seven station ARIANNA test bed deployed at the Moore's Bay site. Only those stations based on the SST data acquisition hardware will be considered here, with the previous generation of detectors being independently analyzed in \cite{2015ARIANNALimitsPaper}. Data collected between December \nth{8}, 2014 and February \nth{5}, 2019 will be analyzed.

\subsection{Operational Livetime}
\label{sec:Livetime}

\begin{figure}
  \centering
    \includegraphics[width=0.5\textwidth]{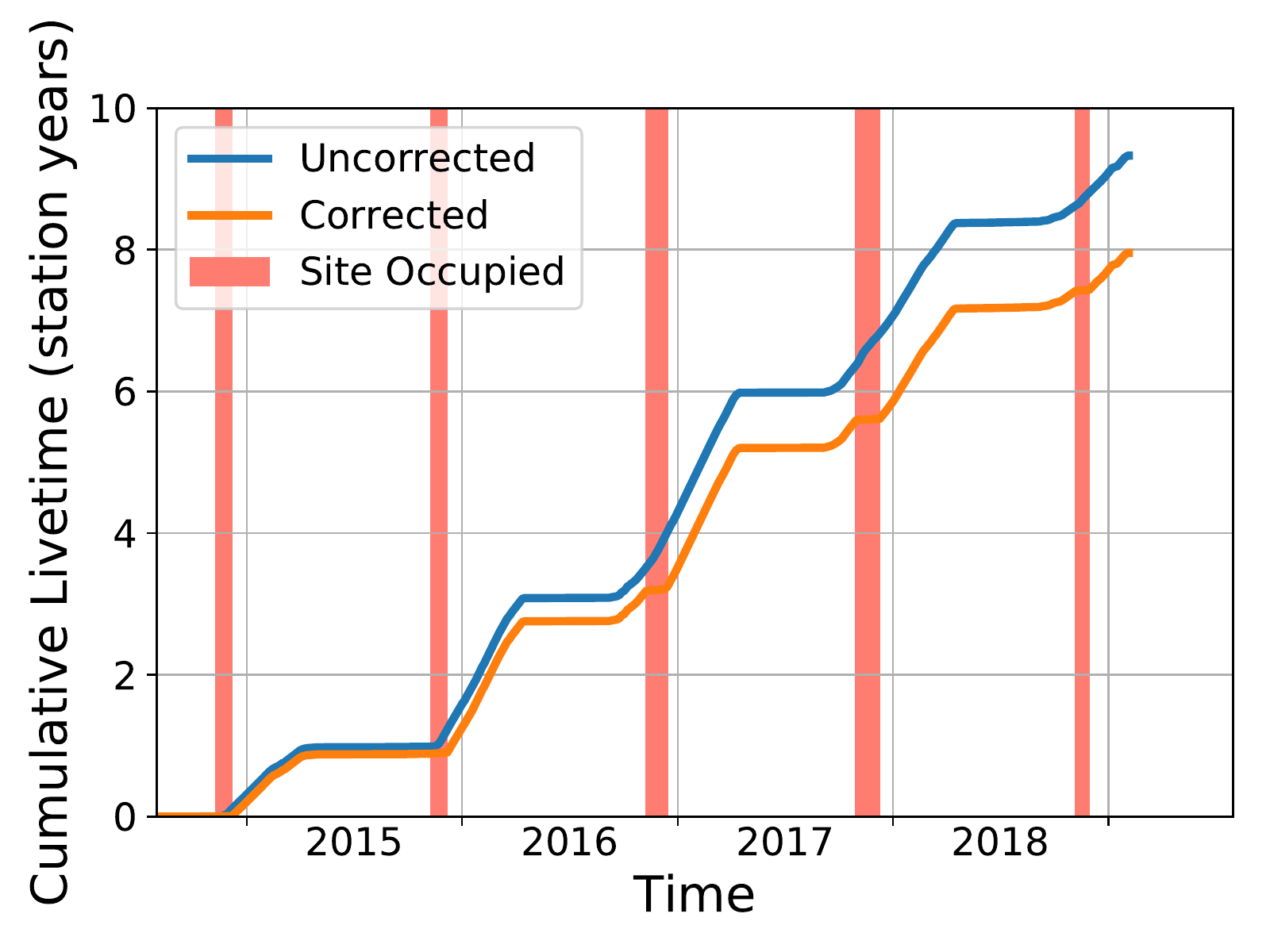}
    \includegraphics[width=0.47\textwidth]{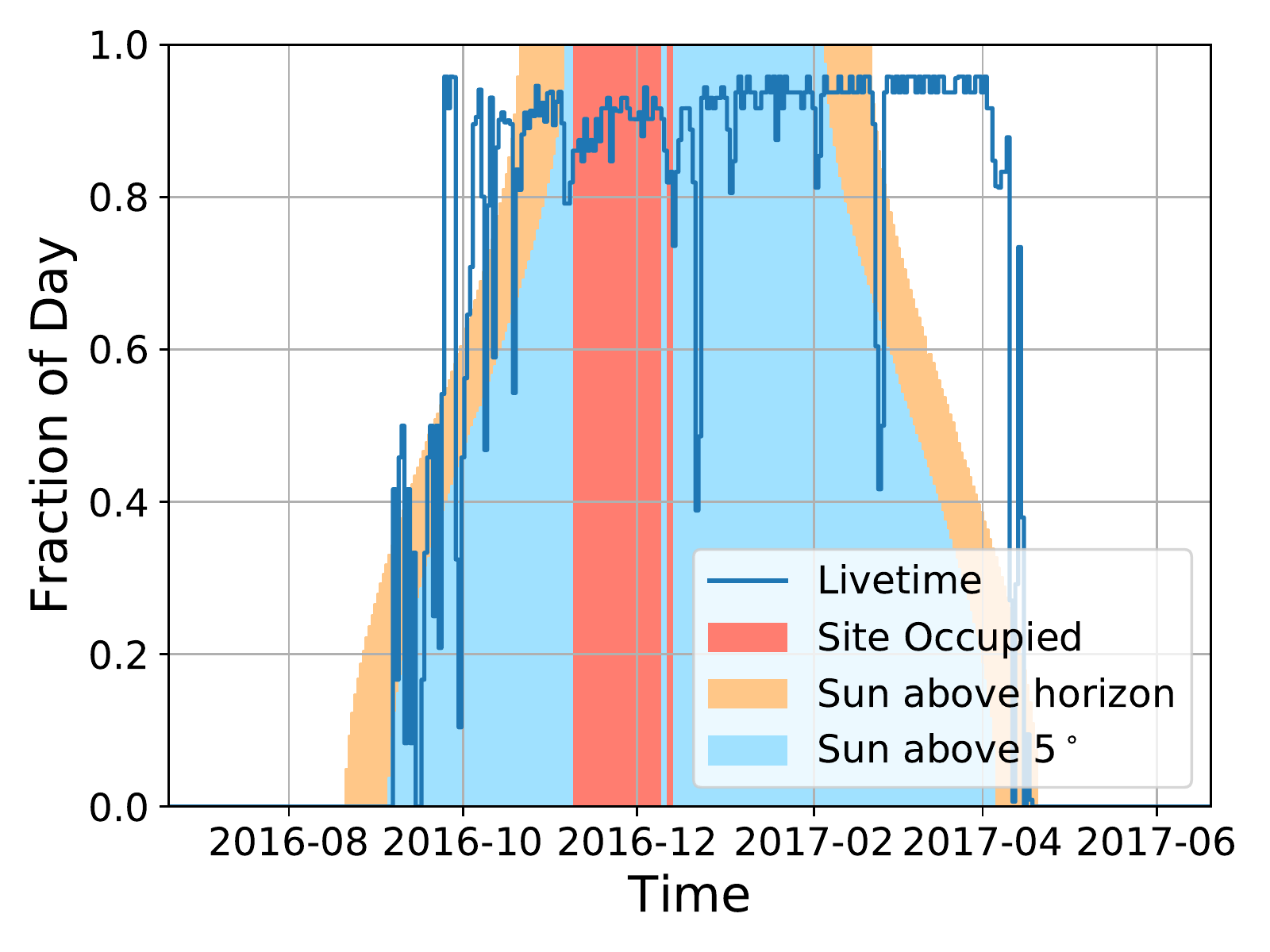}
  \caption{(left) Cumulative livetime for all SST based neutrino stations since deployment.  The corrected curve represents the livetime used for analysis (details in Sec.~\ref{sec:Livetime}). Since the stations rely on solar power, they shut down during the winter and do not accrue livetime. (right) The livetime fraction per day for a solar-powered station in Moore's Bay.  Configurations are tuned to maximize livetime after the end of the field season, after which stations typically show livetimes $> 90\%$. Shaded regions show the fraction of each day in which the sun is above the horizon, and greater than $5^\circ$ above the horizon.}
  \label{fig:HRALivetime}
\end{figure}

Current ARIANNA hardware does not take data during communication with the server in the US. This introduces downtime which is dependent on the station configuration.  Since ARIANNA is a test array, station configurations are often not set to optimize for livetime.  During the 2016-2017 season, however, there was a concerted effort to run the array in an efficient manner.  Stations were shown to reliably operate with 90\% livetime efficiency for extended periods (see Fig.~\ref{fig:HRALivetime}), collecting an average of 151 days of livetime per station through the season.

The livetime calculated for analysis excludes times when the field camp at the ARIANNA site was occupied as a precaution against contaminating the data due to camp activities. In principle, only small periods of time involving pulsed radio studies or operating equipment near stations should be unsuitable for analysis, so this is a conservative measure of the usable livetime. Several hours were also removed on Dec 11-\nth{13}, 2016 during the operation or ANITA's HiCal pulser \cite{PROHIRA201960}, which was observed by the stations in Moore's Bay.

The livetime is also corrected for DAq deadtime due to event readout, which only becomes significant during infrequent periods where events rates increase above \SI{1}{Hz}.  After these corrections, the stations collected a total of 2906.9 days (7.96 station years, with a 365 day year) of livetime for analysis in the period between December 2014 and February 2019 (see Fig.~\ref{fig:HRALivetime}), which is the time frame for this analysis. This figure is based on the data which has been successfully transferred from the ARIANNA stations.  We estimate that the uncollected data, which can be physically recovered from the stations' SD cards, would increase the livetime for analysis by approximately 5\%.

\subsection{Monte Carlo Simulations}
\label{sec:MonteCarlo}

The \emph{ShelfMC} Monte Carlo code \cite{ChrisThesis} was used to calculate the sensitivity of the detector, as well as to generate parameter distributions for the neutrino signal space.  \emph{ShelfMC} is based on a fork of the icemc simulation originally developed by the ANITA collaboration \cite{Cremonesi:2019zzc}.  This section provides a brief overview of the simulation method.

Neutrino interaction vertices are generated uniformly within a fiducial volume, with isotropically distributed arrival directions. Neutrino flavor is also uniformly and randomly assigned, with no special treatment given between $\nu$ or $\bar{\nu}$.  The frequency domain Askaryan signal near the vertex is calculated according to the parametrization in \cite{AlvarezMuiz2000}, which is validated against results from the ZHS Monte Carlo \cite{Zas1992}.  The signal is then propagated through the ice, via a direct and reflected path.  The signal strength after accounting for antenna response is calculated for each channel of the detector, and is counted as a triggered event if it satisfies a 2-of-4 majority logic trigger above a specified threshold.  

Each trigger is assigned a weight value, which is the probability that the neutrino reached the simulation volume (and was not absorbed by the Earth) to interact at the vertex location. The event weight also contains a factor of $\rho(z)/\rho_{ice}$ where $\rho(z)$ is the ice density at the interaction vertex, and $\rho_{ice}=\SI{0.9167}{g cm^{-3}}$ is the density of solid ice. This factor accounts for the variation in the density of nuclear targets within the ice shelf.  Weight values are further modified to account for effects of $\nu_{\tau}$ regeneration, wherein $\tau$ produced by the charged current interaction of $\nu_\tau$ quickly decay to produce a $\nu_\tau$ at lower energy (and longer interaction length) en route to the detector, as described in \cite{Halzen1998}.

\begin{figure}
  \begin{centering}
    \includegraphics[width=0.8\textwidth]{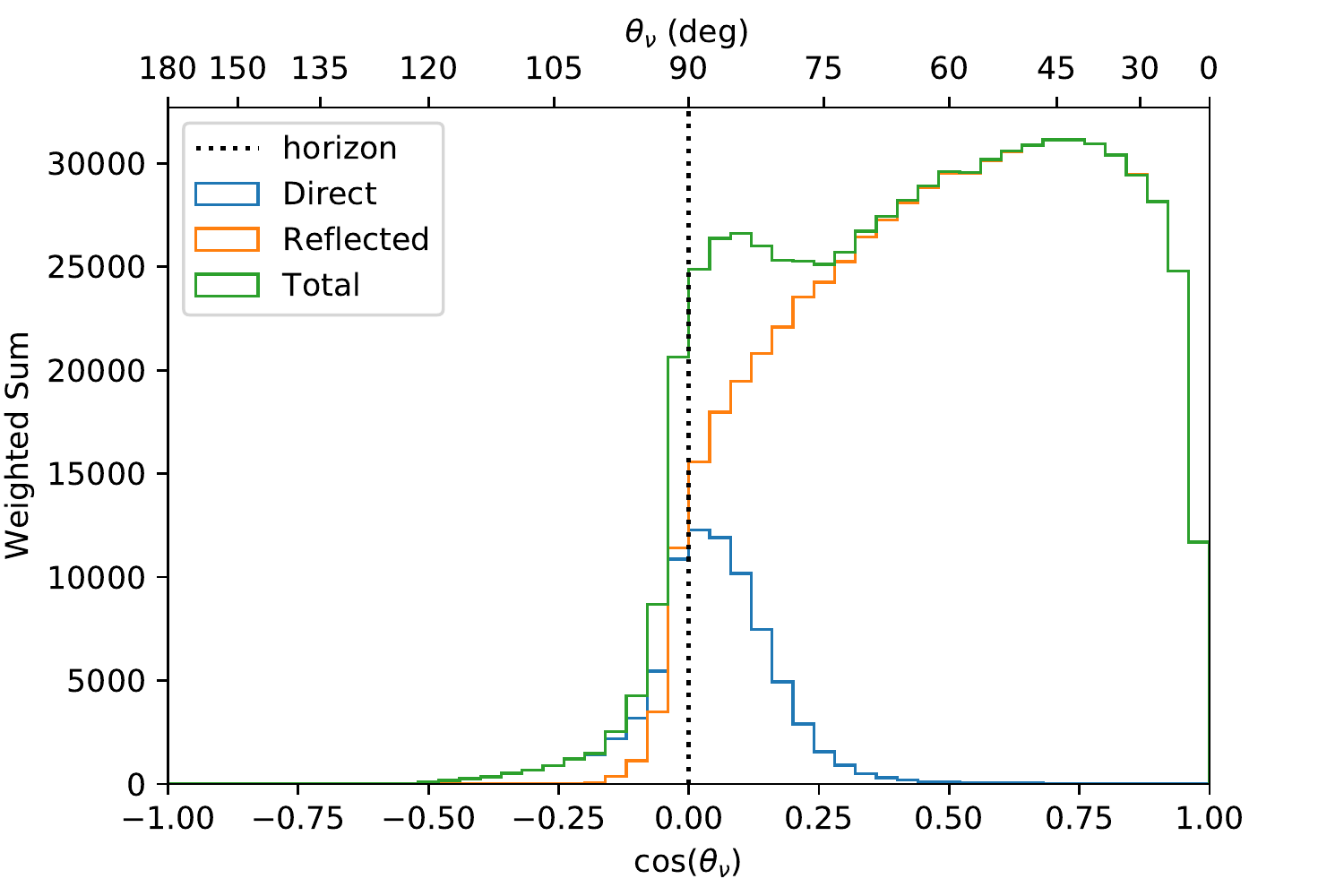}
    \caption{Arrival zenith angle for neutrinos which lead to a trigger in simulation. $0^{\circ}$ represents a downward going neutrino. The addition of the reflected signal greatly amplifies the effective volume, while also generating more uniform sky coverage. See Table\ \ref{tab:SimulationParamsAsBuilt} for simulation parameters. Neutrino energies are distributed according to a GZK spectrum, identically to the neutrino signal space in Sec.~\ref{sec:SigSpace}. }
    \label{fig:4SigSignalThetaNu}
  \end{centering}
\end{figure}

The Askaryan signal amplitude falls rapidly off the Cherenkov cone, especially at high frequencies. This limits the arrival direction for triggering neutrinos such that the signal path to the detector must lie close to the Cherenkov cone.  This constrains the arrival direction to be near horizontal for direct signals, but the reflection at the ice/sea boundary allows for neutrinos from the entire sky to potentially trigger.  This effect is illustrated in Fig.~\ref{fig:4SigSignalThetaNu}. 

\begin{figure}
  \centering
  \includegraphics[width=0.9\textwidth]{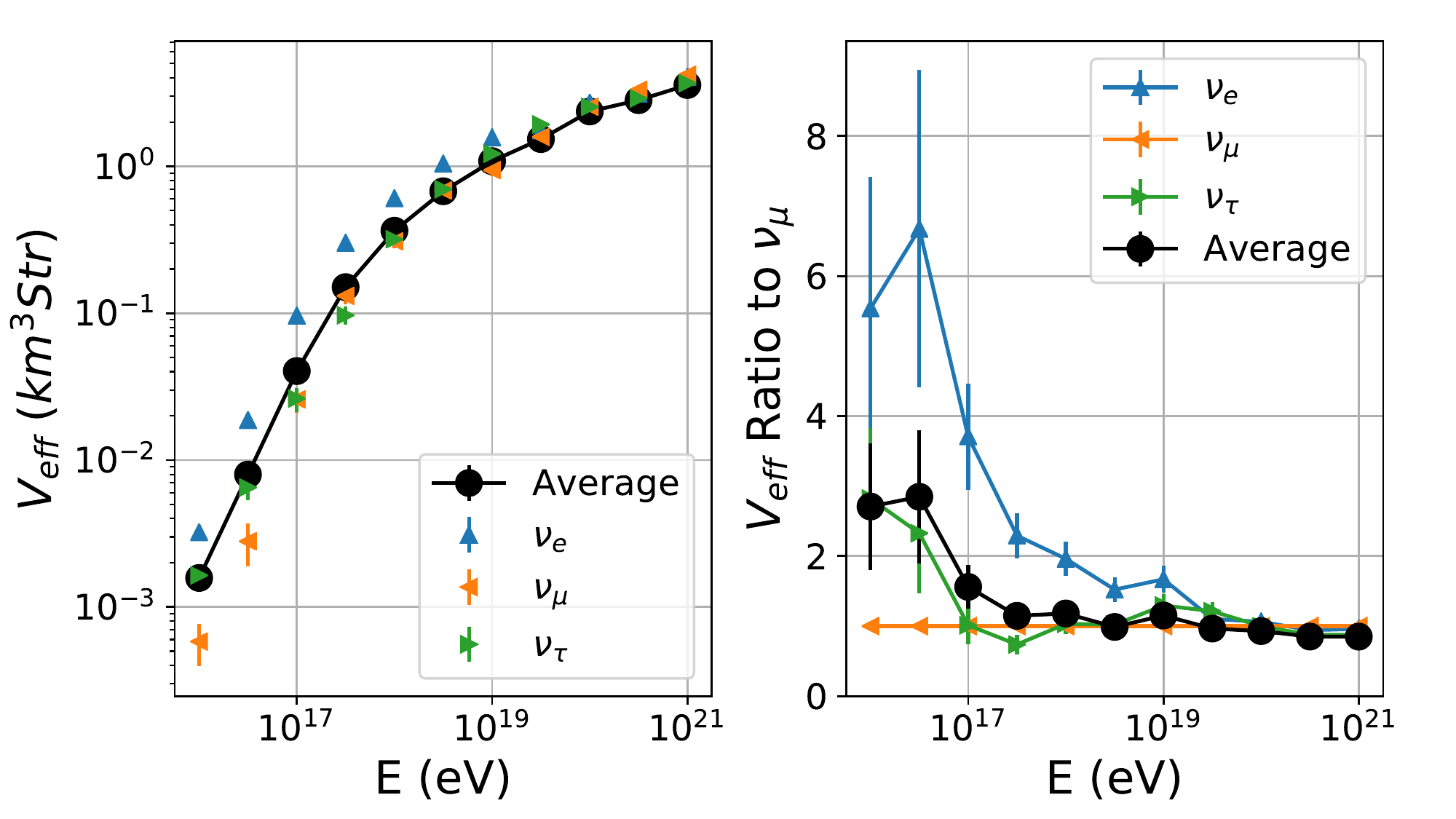}
  \caption{(left) Effective Volume for a single 4-channel station, simulated according to the simulation parameters in \ref{tab:SimulationParamsAsBuilt}. (right) Ratio of single-flavor $V_{eff}$ to the $\nu_\mu$ effective volume.  $\nu_e$ effective volume is enhanced due to the presence of an electron (or positron) induced electromagnetic shower for charged current interactions, though this advantage diminishes at higher energy due to the LPM effect. Low energy $\nu_\tau$ see an enhancement due to tau regeneration effects, with a GZK spectrum assumed \cite{Engel2001}. }
  \label{fig:ShelfMCStandardVeff}
\end{figure}

Electromagnetic showers resulting from $\nu_e$ charged current interactions are elongated due to the LPM effect, which is treated in the simulation as a narrowing of the Cherenkov cone width, according to \cite{AlvarezMuiz1998}. It is expected that this picture is complicated at shower energies above \SI{e18}{eV} because the electromagnetic component breaks into multiple sub-showers (see \cite{Gerhardt2010} and references therein).  In this work, simulated elongation of the EM shower effectively eliminates their contribution to the effective volume above \SI{e19}{eV} (see Fig.~\ref{fig:ShelfMCStandardVeff}).  This treatment avoids the complication of event-by-event variations in the highest energy EM showers, leading to an underestimation of the effective volume for charge current $\nu_e$ interactions at the highest energies.

The ice shelf is modeled as having an exponential density profile \cite{hanson2015,ForbiddenPropagation2018}, which leads to the following dependence of index of refraction with depth:
\begin{equation}
  n(d) = n_{ice} + (n_{0}-n_{ice})e^{-d/C} \quad \textrm{where} \quad d > 0,\: n_0=1.3,\: n_{ice}=1.78,\: \textrm{and}\: C=\SI{34.48}{m}.
  \label{eqn:FirnIndex}
\end{equation}

As the index of refraction changes with depth, it causes the signal ray path to bend, and creates a region in which signal can not propagate to the detector, which is referred to here as the "shadow zone."  Recent measurements show that it is, in fact, possible to measure signals generated within this zone \cite{ForbiddenPropagation2018}. However, only a small fraction of the Askaryan signal couples into such a propagation mode such that the influence of \emph{shadow zone signals} on the effective volume is small. An exception might be very high energy events where sufficient energy fluence is channeled into this propagation mode \cite{LahmannICRC2019}. For this analysis, no interaction vertices are generated in the shadow zone.
In the shelf-mc code, the shadow zone boundary is calculated via a numerical ray tracing technique using the density profile from Eq.~\eqref{eqn:FirnIndex}. For emitter positions outside the shadow zone, the signal ray path is approximated with two straight line segments where the ray is refracted once at a depth corresponding with the bottom of the firn (for more details see \cite{ChrisThesis}).

\begin{table}
  \centering
  \begin{tabular}{c|c}
    Parameter & Setting \\
    \hline
    Spectrum & GZK \\
    Energy Range & $10^{15.5}$ - \SI{e21.5}{eV} \\
    Ice Thickness & \SI{575}{m} \\
    C (see Eqn.\ \ref{eqn:FirnIndex}) & \SI{34.48}{m} \\
    Firn Depth & \SI{68.96}{m} \\
    Depth Averaged Attenuation Length & \SI{500}{m} \\
    Noise Temperature & \SI{350}{K} \\
    Bandwidth & 50 - \SI{1000}{MHz} \\
    Noise Before Trigger & Disabled \\
    Reflection Coefficient & 0.9 \\
    Trigger Threshold & $4\sigma$ \\
    Majority Logic & 2 of 4 \\
    Tau Regeneration & Enabled \\
    Shadowing & Enabled \\
  \end{tabular}
  \caption{\emph{ShelfMC} simulation parameters for the currently deployed ARIANNA stations. While there is no discrete transition between the firn and the bulk ice, we take the firn depth to correspond to 2 e-foldings of density ($depth = 2C$). A detailed description of each parameter can be found in \cite{ChrisThesis}.} 
  \label{tab:SimulationParamsAsBuilt}
\end{table}

For effective volume simulations, neutrinos at several energies are simulated, and the weighted number of triggers is used to calculate the water-equivalent effective volume according to
\begin{equation}
\centering
V_{eff} = V_{fid} \cdot \frac{\rho_{ice}}{\rho_{water}} \cdot 4\pi \cdot \frac{1}{n} \cdot \sum_{i=1}^{n_{trig}} w_i,
\label{eqn:VeffEquation}
\end{equation}
where $V_{fid}$ is the fiducial volume, $n$ is the total number of simulated events, and $w_i$ are the weights of the triggered neutrinos. The normalization factor of $4\pi$ corresponds to an integration over the solid angle of the simulation. This is a typical choice for expressing the sensitivity to a diffuse/isotropic flux, which is the primary interest of this analysis. The effective volume for a single ARIANNA test bed station is shown in Fig.~\ref{fig:ShelfMCStandardVeff}.

\subsection{Neutrino Signal Simulation}
\label{sec:SigSpace}

In order to model the sensitivity of the ARIANNA detector, it is first necessary to generate a population of simulated neutrino signals from Monte Carlo.  For this purpose, one billion neutrino interactions were simulated in \emph{ShelfMC} with a GZK energy spectrum from \cite{Engel2001}, using a detector configuration matching the currently deployed ARIANNA test bed stations in Moore's Bay (see Table \ref{tab:SimulationParamsAsBuilt} for details).  This produced a sample of approximately 1.3 million events which triggered the simulated station at a $4\sigma$ level above the thermal noise $V_{rms}$.

\begin{figure}
  \centering
    \includegraphics[width=0.49\textwidth]{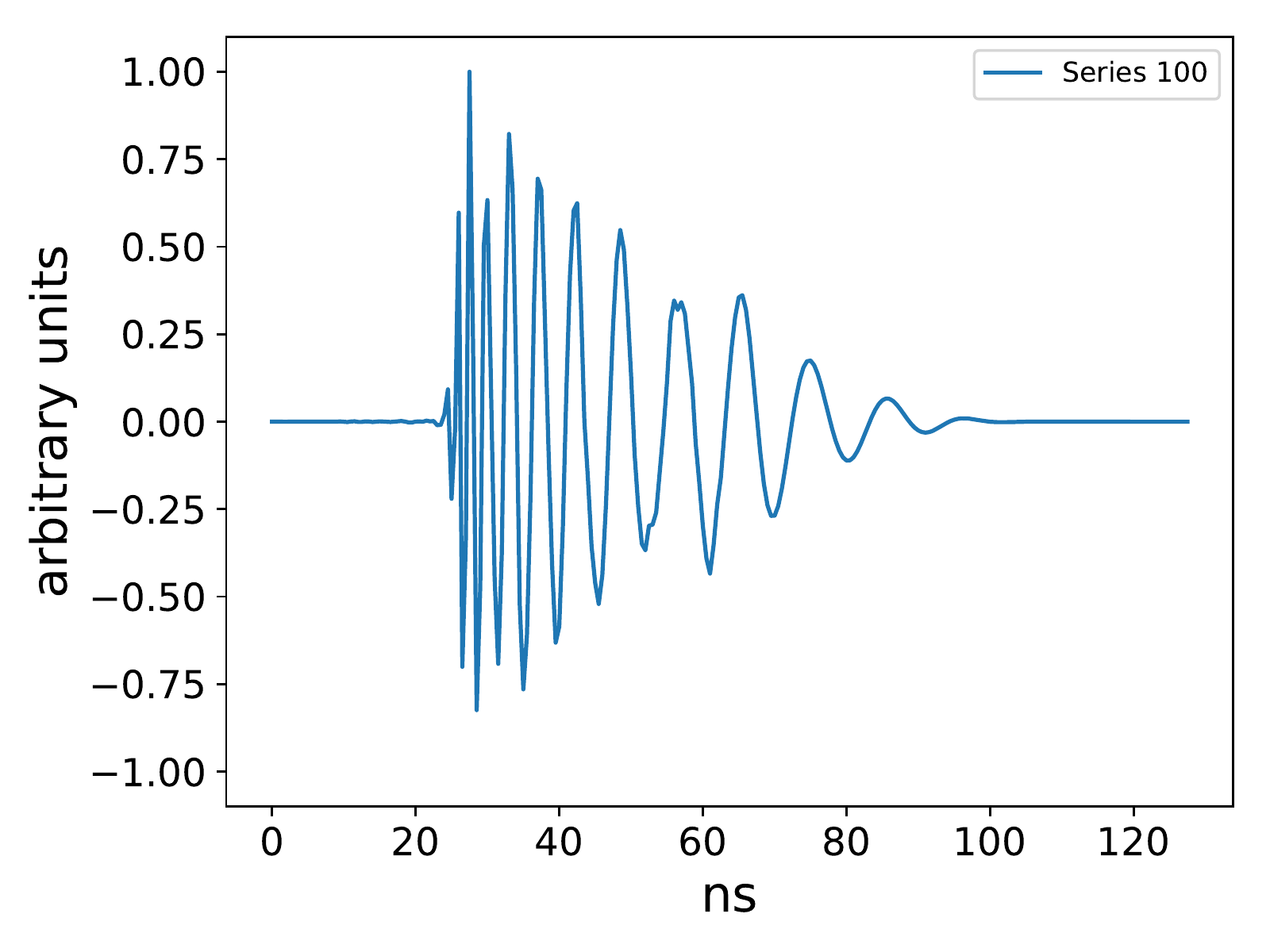}
    \includegraphics[width=0.49\textwidth]{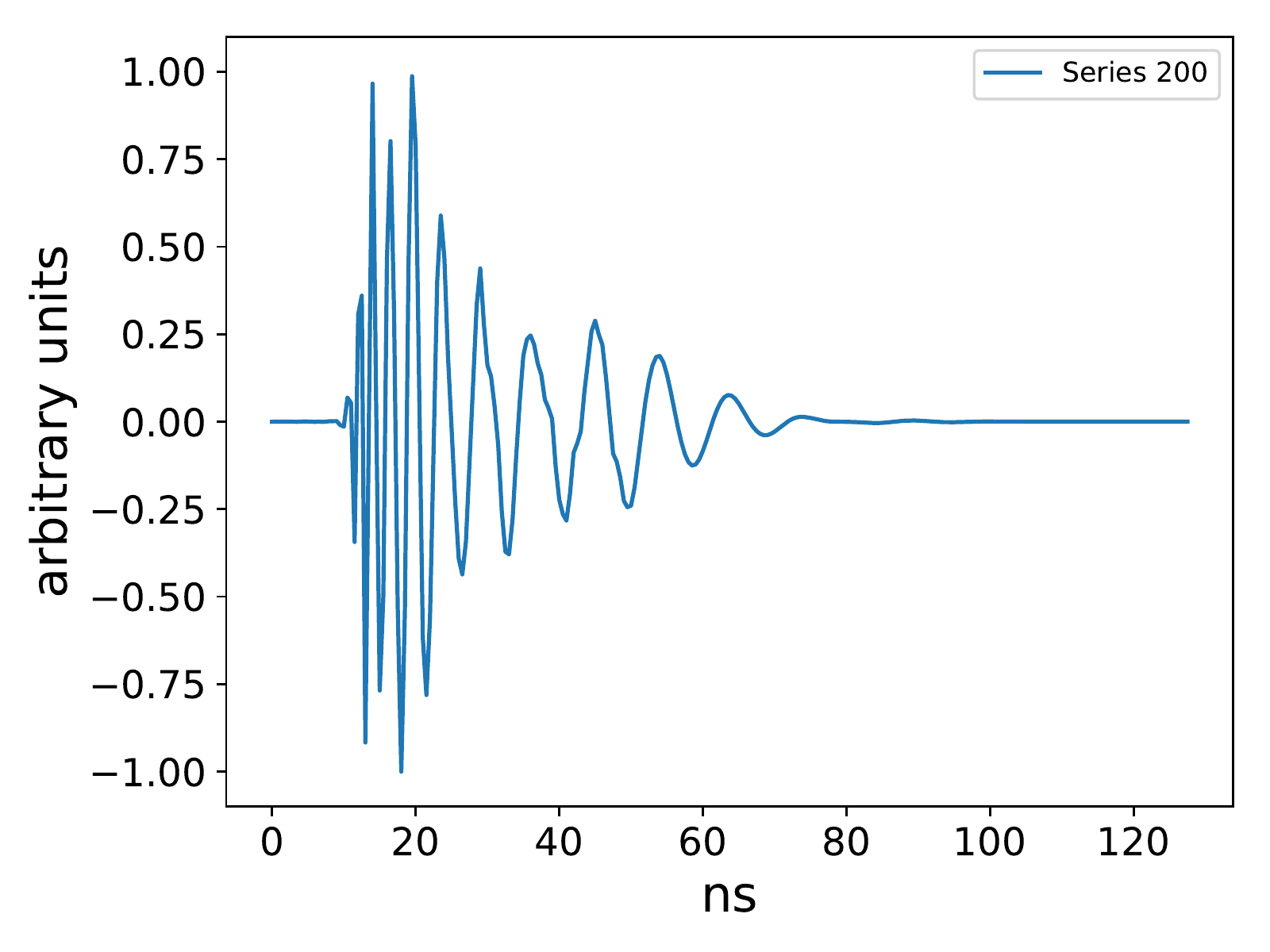}
  \caption{Reference neutrino templates for 100 series (left) and 200 series (right) amps.  Both templates correspond to on-cone neutrino signals with E-plane and H-plane angles of $30^\circ$ for the LPDA response.  The chirped response, with the higher frequency components arriving earlier in the pulse is due to both the antenna dispersion and amplifier group delay, and is characteristic of the system response to an impulse.}
  \label{fig:RefTemplates}
\end{figure}

To produce the time domain signal for neutrino events, a template procedure is used. Neutrino templates are generated which incorporate the direction dependent response of the LPDA's as well as the amplifier response, according to the procedure in \cite{2015ARIANNATimeResponsePaper}.  Templates are spaced in $10^\circ$ increments in E-plane and H-plane angle of the antenna (see Fig.~1 of \cite{2015ARIANNATimeResponsePaper} for a definition of the E and H plane of the LPDA), and $0.5^\circ$ increments in viewing angle. Two versions of the ARIANNA amplifiers are in use in the pilot array (see Sec.~\ref{sec:amps}), which have slight differences in their impulse response.  To account for these differences, separate templates are constructed for each type of amplifier (see Fig.~\ref{fig:RefTemplates} for example templates), and the stations are analyzed separately. For each triggered \emph{ShelfMC} event, the arrival direction and polarization relative to the antenna orientation, and the viewing angle relative to the Cherenkov cone are used to choose the most appropriate template.  The template is then scaled according to the appropriate amplitude, and finite bandwidth random noise is added.  This data is converted to the same format as typical ARIANNA data, and run through exactly the same analysis as the triggered events.

\section{Analysis}
\label{sec:analysis}
A search for neutrino signal was carried out using all available data as described in Sec.~\ref{sec:data}. The methods and results from this analysis are discussed in detail in this section.

\subsection{Definition of Analysis Variables}

The main technique for the separation of signal and background in this analysis is a template matching procedure, similar to that which was carried out in \cite{2015ARIANNALimitsPaper}. The signal on each channel is compared to a specific reference template, shown in Fig.~\ref{fig:RefTemplates}. The signal is shifted in time relative to the template until the absolute value of the Pearson correlation coefficient is maximized. This maximum value, $\chi$, is a measure of similarity between the signal and reference template, with values ranging from 0 (no correlation) to 1 (identically similar). After calculating $\chi$ for each channel, we combine the results into a single parameter per event, $\chi_{ave}$, which is the greater of the two averages between co-polarized pairs of LPDA's.  High values of $\chi_{ave}$ essentially imposes a condition that pairs of antennas see similar signals, which is expected, as the neutrino signal does not significantly change over the \SI{6}{m} separation of the antennas.

By also considering the signal amplitude, it is possible to leverage the general behavior that a high signal-to-noise ratio (SNR) neutrino signal event is likely to have a larger value of $\chi_{ave}$.  This greater discriminating power for large amplitude events raises the analysis efficiency. In this text the SNR of an event will be defined as $V_{PTP} / 2\sigma$, where $V_{PTP}$ is the peak-to-peak amplitude and $\sigma$ is the root-mean-squared voltage of minimum bias noise events.

\begin{figure}
  \begin{centering}
        \includegraphics[width=0.49\textwidth]{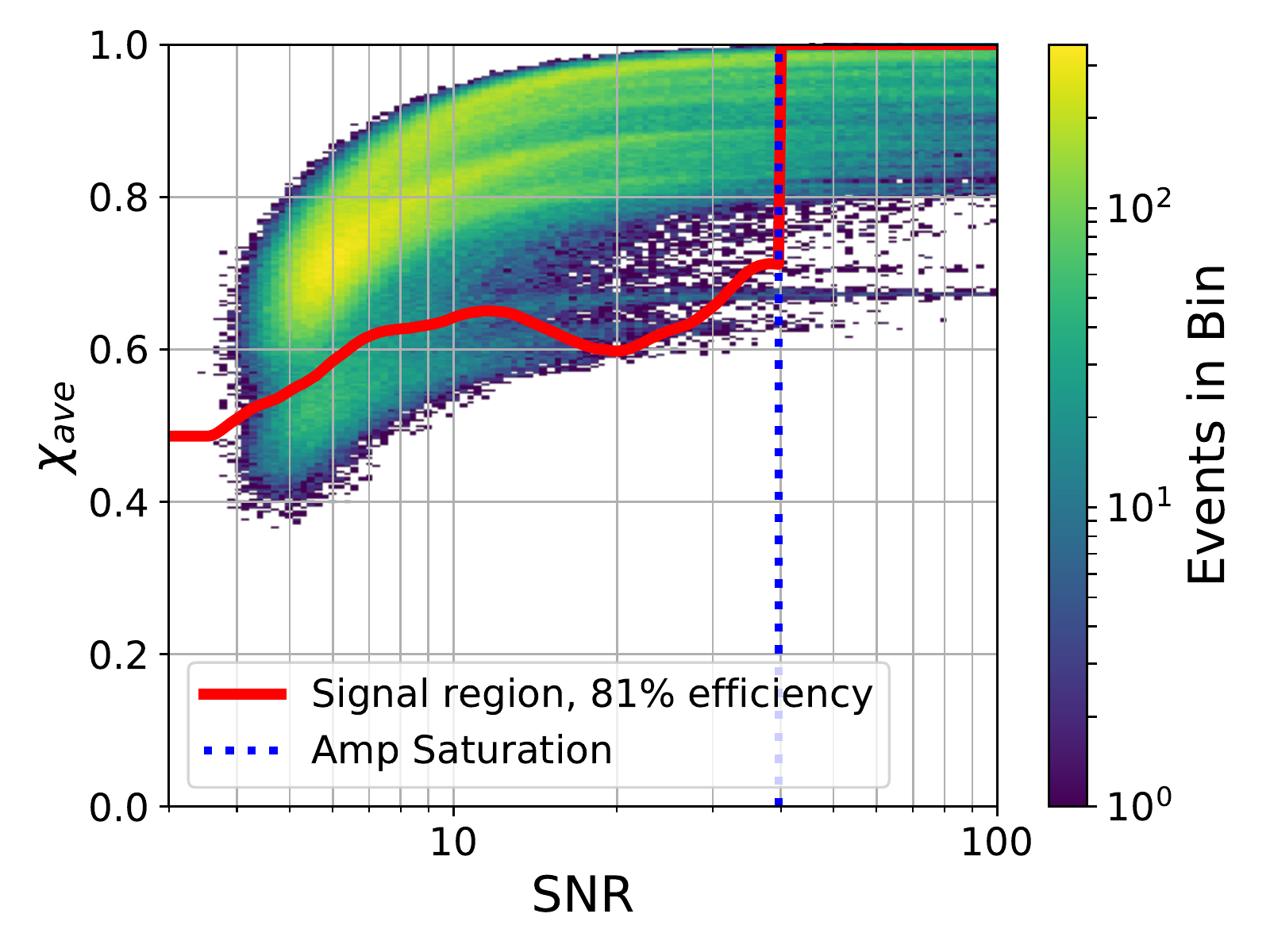}
        \includegraphics[width=0.49\textwidth]{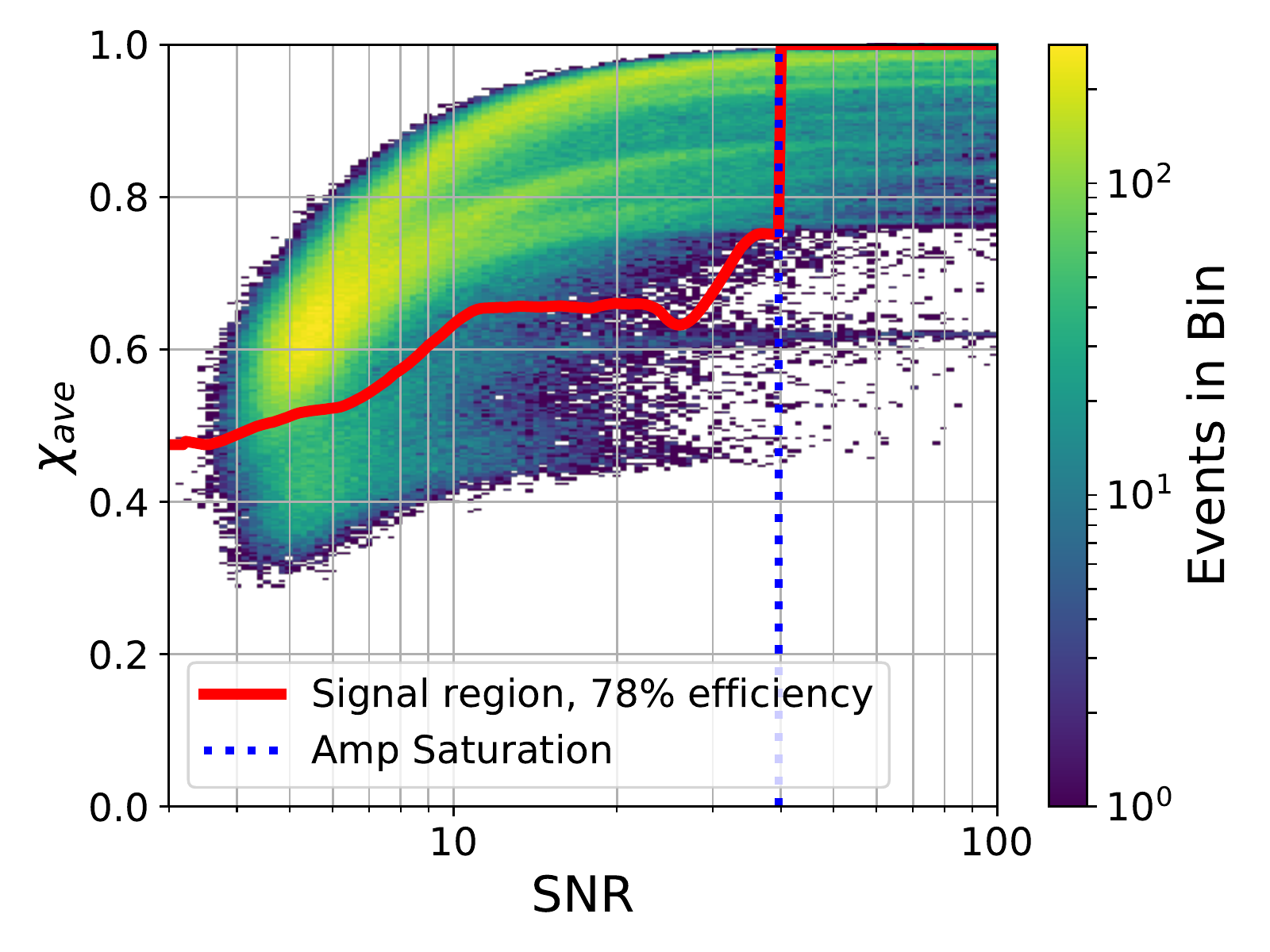}
    \caption{Distribution of simulated neutrino signal in $\chi_{ave}$ vs SNR. Different neutrino templates are used for stations with Series 100 amps (left) and Series 200 amps (right). Larger values of SNR correspond to larger values in $\chi_{ave}$, since lower SNR leads to worse cross-correlation to the reference template. The dashed blue line represents a \SI{800}{mV} cutoff where the amplifiers would be strongly non-linear. The red line represents the lower bound of the signal region, which is discussed in Sec.~\ref{sec:SignalCut}.}
    \label{fig:SignalDistAnalysis}
  \end{centering}
\end{figure}

We create a 2D parameter space using $\chi_{ave}$ and SNR in which we will define a signal region for this analysis.  Analysis in a space of template correlation vs amplitude has previously been successfully used to identify cosmic-ray air showers in ARIANNA stations \cite{NellesCRs2017}. The distribution of the simulated neutrino signal in this space is shown in Fig.~\ref{fig:SignalDistAnalysis}.

\subsection{Defining the Signal Region}
\label{sec:SignalCut}

\begin{figure}
  \begin{centering}
        \includegraphics[width=0.49\textwidth]{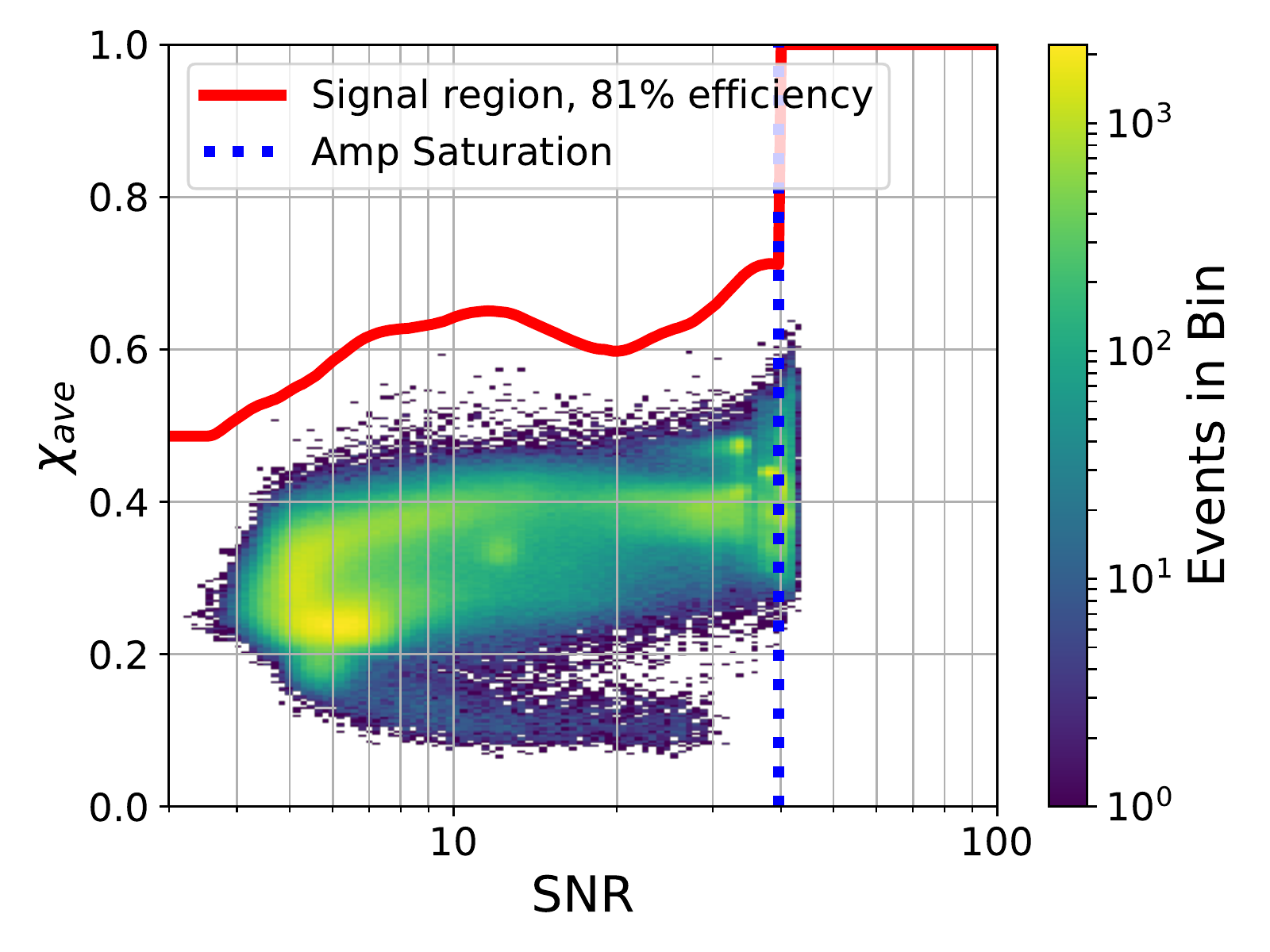}
        \includegraphics[width=0.49\textwidth]{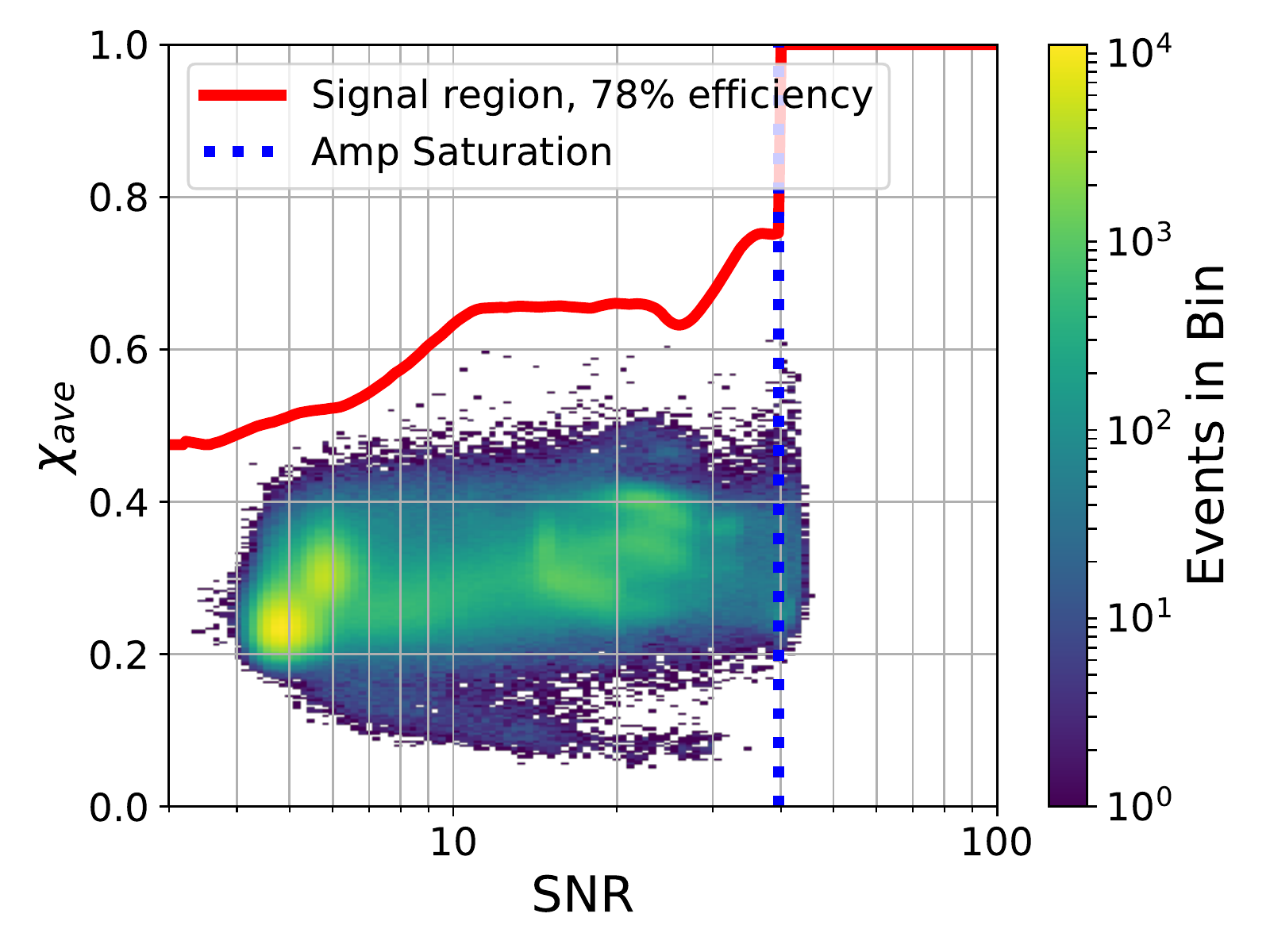}
    \caption{Distribution of triggered events in $\chi_{ave}$ vs SNR for series 100 amplifiers (left), and series 200 amplifiers (right).  Larger values of SNR correspond to larger values in $\chi_{ave}$ up to $SNR \approx 10$, since lower SNR leads to worse cross-correlation to the reference template. The blue horizontal line represents a \SI{800}{mV} where a noticeable pile-up effect is observed due to amp clipping. The red line represents the lower bound of the signal region, which is discussed in Sec.~\ref{sec:SignalCut}.  No triggered events remain in the signal region.}
    \label{fig:TriggeredDistAnalysis_100}
  \end{centering}
\end{figure}

We present the ARIANNA test bed data (cf. Sec.~\ref{sec:data}) in the $\chi_{ave}$ and SNR space in Fig.~\ref{fig:TriggeredDistAnalysis_100}. The data itself is used to define a signal region that maximizes the analysis power, i.e., achieves a high neutrino signal efficiency while rejecting background events efficiently.

\begin{figure}
  \centering
  \includegraphics[width=0.7\textwidth]{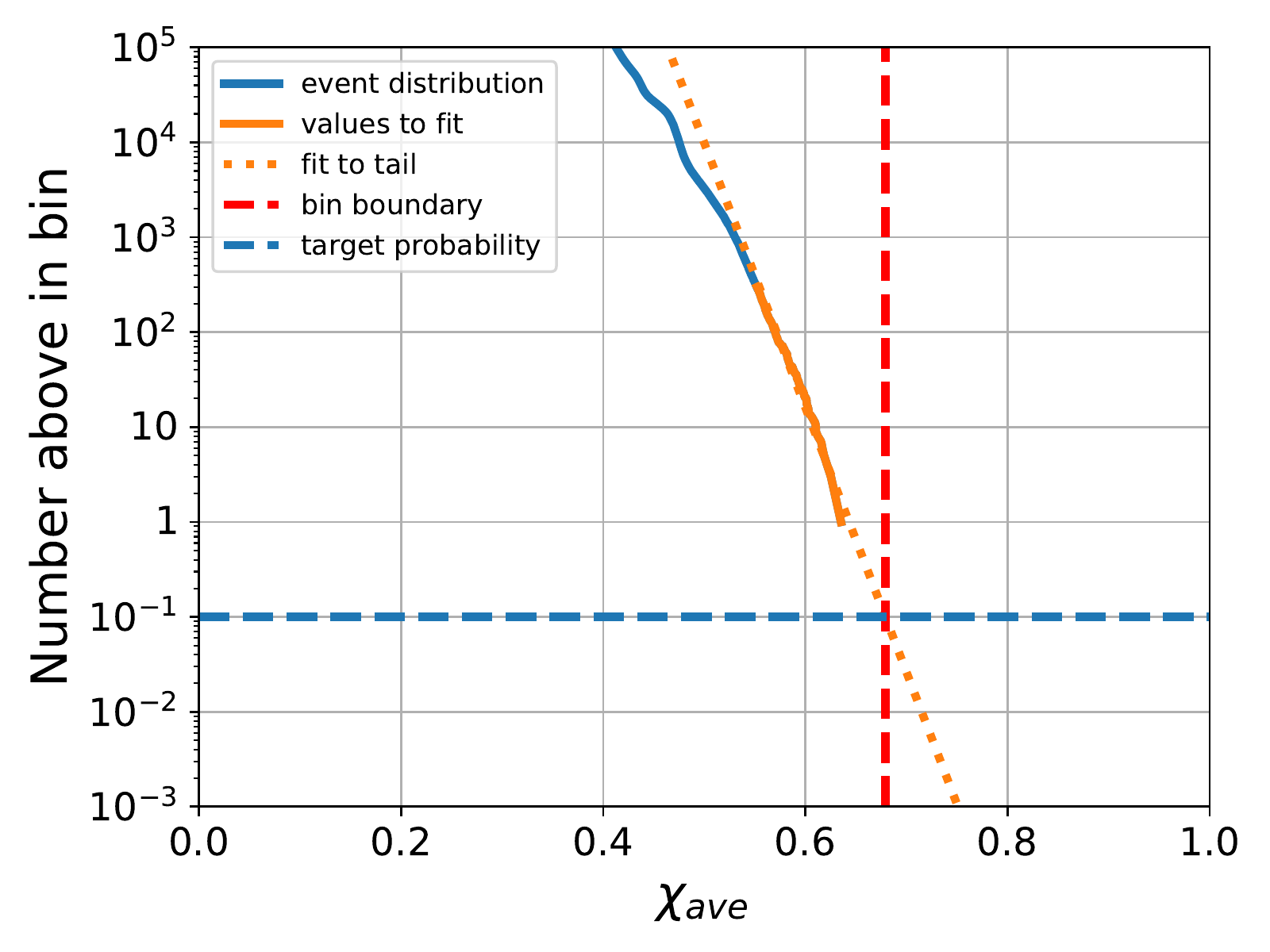}
  \caption{An illustration of the procedure for calculating the signal region boundary for a particular sliding amplitude bin. A power law is fit to the 300 highest $\chi_{ave}$ events in the bin (the single highest value event is excluded from the fit in order to limit the effect of outliers), and extrapolated to intersect the desired background event probability, yielding the signal region lower bound value of $\chi_{ave}$. }
  \label{fig:FitExample}
\end{figure}

We can safely assume that the vast majority of all triggered events are background, and therefore use the data to estimate the background distribution in our 2D parameters space $\chi_{ave}$ and SNR. We bin the data in SNR bins of $\log_{10}\mathrm{SNR} = 0.2$. For each bin we calculate the cumulative histogram in $\chi_{ave}$, i.e., calculate how many events are present with $\chi_{ave}$ greater than or equal to a certain value. For illustration purposes we show one of these distributions in Fig.~\ref{fig:FitExample}. The tail of the cumulative distribution, except for the single highest value, is fit to a power law, and extrapolated to calculate the expected number of background events at higher $\chi_{ave}$ thresholds, extending a technique employed in \cite{2015ARIANNALimitsPaper}.

We calculate the $\chi_{ave}$ thresholds for each SNR bin that result in an expectation of a total of 0.5 background events that pass the signal cut over the total livetime. We treated both amplifier types separately with the expected number of background events for each set weighted by the fraction of the total livetime contribution by that station type. We weighted each SNR bin by the fraction of expected neutrino events in the bin (from Fig.~\ref{fig:SignalDistAnalysis}). The resulting curve is then smoothed by a second order Savitzky-Golay filter \cite{Savitzky1964} and shown as red curve in Fig.~\ref{fig:SignalDistAnalysis} and \ref{fig:TriggeredDistAnalysis_100}. This procedure effectively produces a curve of constant signal to background ratio.

The signal region defined through this procedure contains no triggered events on any ARIANNA station during the complete run, so we conclude that no neutrinos were observed (Fig.~\ref{fig:TriggeredDistAnalysis_100}).  The analysis efficiency of the signal region, which is the fraction of weighted simulated neutrinos which fall above the cutoff is 81\% for the 100 series stations and 78\% for the 200 series. 

\subsection{Limits on the Diffuse UHE Neutrino Flux}
\label{sec:FluxLim}

\begin{figure}
  \begin{centering}
    \includegraphics[width=0.7\textwidth]{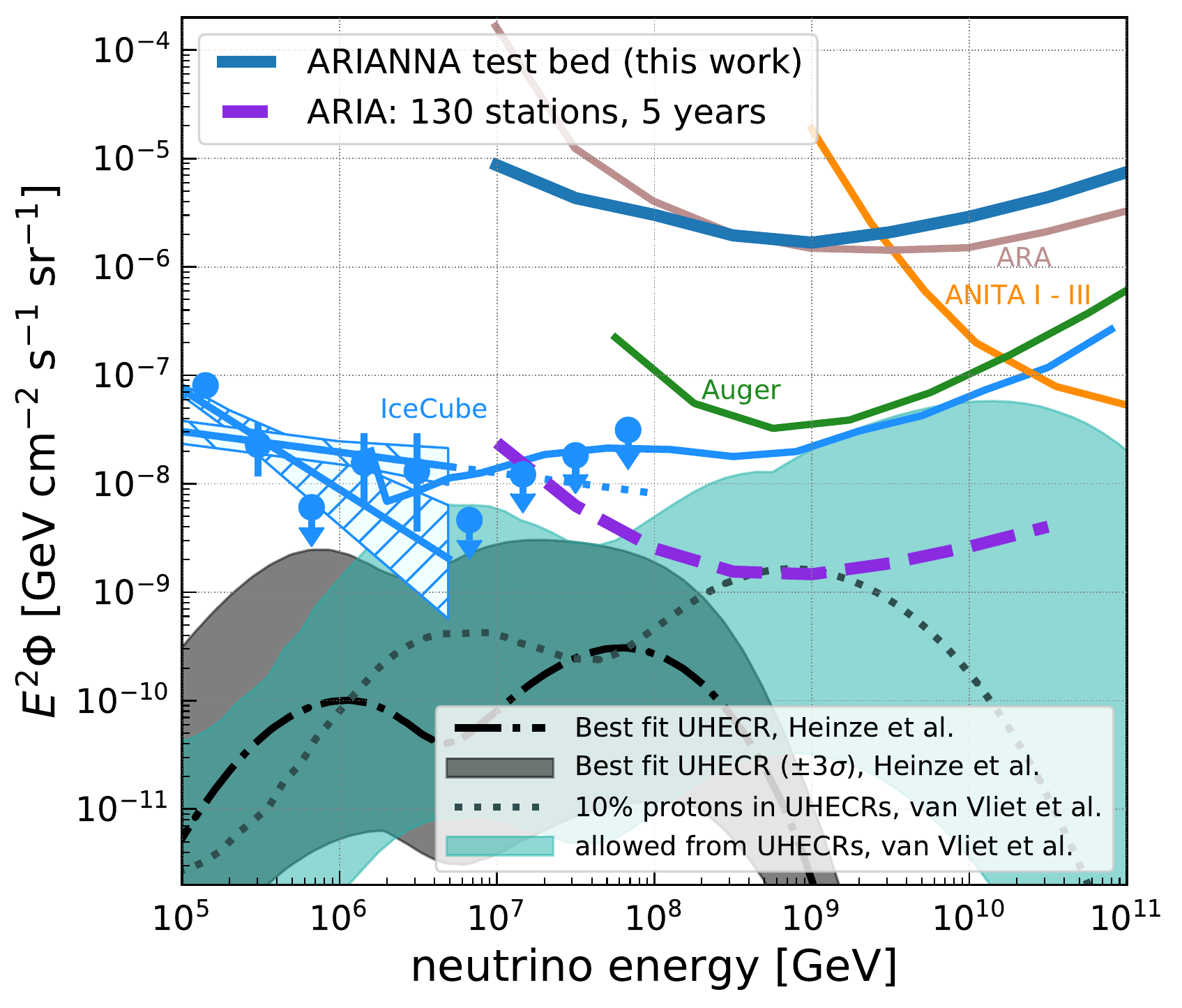}
    \caption{Model independent diffuse flux limit for the ARIANNA test bed for a sliding decade wide energy bin. IceCube measured fluxes of astrophysical neutrino spectra were taken from \cite{kopper2017observation,IcecubeMuonFlux2017} and the limit from \cite{IceCubeLimit2018}. Also shown are published limits from the Pierre Auger \cite{AugerLimit2015} and ANITA collaborations \cite{Gorham2018}.  The published ARA limit from \cite{ARALimit2016} is shown in comparison. To illustrate the neutrino parameter space, diffuse flux models from \cite{Vliet2019} and \cite{Heinze2019} are shown. For details of the calculation see Sec.~\ref{sec:FluxLim}.  The sensitivity for ARIA \cite{COSPAR2019} represents a proposal for an optimized future surface detector based on the ARIANNA technology.}
    \label{fig:MBUpperLimits}
  \end{centering}
\end{figure}

In the absence of observed events in the signal region, a model independent 90\% confidence upper limit on the diffuse neutrino flux is given by,

\begin{equation}
E^2 \Phi(E) \leq \frac{FC_{90} \ E \ L(E)}{\ln10 \ d\log E \ V_{eff} \ \sum_i  (\epsilon_i \ t_i)}
\label{eqn:FCFLuxLimit}
\end{equation}

where $E$ is the energy of the neutrino, $d\log E=1$ is the bin width (set to 1 decade), $V_{eff}$ is the effective volume of a single ARIANNA station averaged over all flavors, $L$ is the water equivalent neutrino interaction length calculated using the cross section in \cite{Connolly2011}, and $t_i$ and $\epsilon_i$ are, respectively, the total livetime and the analysis efficiency for each ARIANNA station i. $\epsilon_i$ for each station is given by the analysis efficiency for stations with that amplifier type, as calculated in Sec.~\ref{sec:SignalCut}. $t_i$ for each station is the useful livetime collected, according to the conditions outlined in Sec.~\ref{sec:Livetime}.  $FC_{90}=2.44$ is the Feldman Cousins 90\% confidence upper limit for 0 measured events and an expected background of 0 events in each decade-wide energy bin \cite{Feldman1998}. Though the expected background is estimated to be 0.5 events over the entire energy interval, or approximately 0.08 events per decade energy bin, we chose to set the expected number of background events per energy bin to 0.  This procedure produces a slightly larger upper limit than a more accurate calculation that computes the expected background in each energy bin.

The results of this calculation for the seven stations of the ARIANNA test bed are shown in Fig.~\ref{fig:MBUpperLimits}. For a decade wide logarithmic bin centered at a neutrino energy of $10^{18}$ eV, the 90\% confidence upper limit is $E^2\Phi=\SI{1.7e-6}{GeV cm^{-2}s^{-1}sr^{-1}}$. While the diffuse flux limit from the test bed is not competitive with the current state of the art, it represents an order of magnitude improvement over the previous published limit from the ARIANNA collaboration \cite{2015ARIANNALimitsPaper} and demonstrates the long-term reliability of the ARIANNA hardware in-situ.

Also shown in Fig.~\ref{fig:MBUpperLimits} is a projected limit from an optimized array based on ARIANNA technologies (the ARIA proposal \cite{COSPAR2019}). This optimized design achieves greater effective volume per station by lowering thresholds to $3\sigma$ (by a more advanced L1 trigger on the FPGA) and taking advantage of the colder ice at the South Pole.  Assuming that such an array could operate at 100\% uptime on a wired power grid or through a combination of solar and wind power, a 5 year run would be sensitive enough to limit the proton fraction of the highest energies to 10\% or less \cite{COSPAR2019}.

\subsection{Transient Event Sensitivity}

\begin{figure}
  \begin{centering}
    \includegraphics[width=0.6\textwidth]{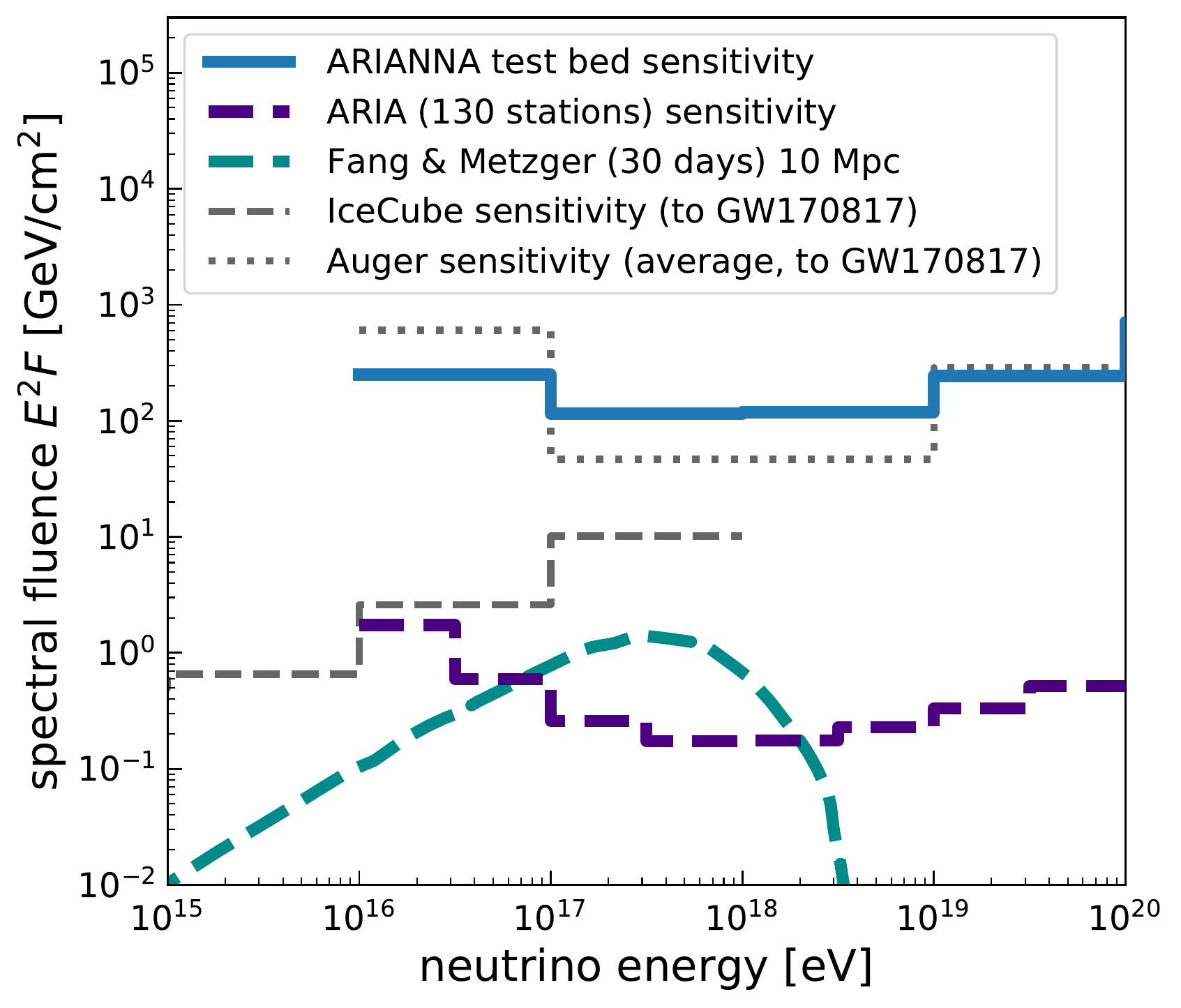}
    \caption{Sensitivity of the ARIANNA test bed to transient events. Sensitivities of other experiments in the direction of the neutron star merger GW170817 (black lines) are taken from \cite{Albert2017}.  A theoretical flux of ultra-high energy neutrinos from a neutron star merger \cite{Fang2017} is also shown (dashed cyan line). The ARIA sensitivity \cite{COSPAR2019} (dashed purple line) represents the capability of a future detector based on the ARIANNA technology to detect transient sources in its field of view.}
    \label{fig:TransientLimit}
  \end{centering}
\end{figure}

As equation \ref{eqn:FCFLuxLimit} shows, the flux sensitivity for diffuse sources depends on effective volume and time, which  benefits from the steady accumulation of livetime over years of operation.  However, the sensitivity to a specific transient point source depends solely on effective volume and whether the object is in the field-of-view.

The transient source sensitivity of the seven station  ARIANNA test bed is fairly uniform over the entire upward sky (Figs.~\ref{fig:4SigSignalThetaNu} and \ref{fig:SkyCoverage}).  It is comparable to the shown limit from the Auger collaboration in the direction of the neutron star merger GW170817 (Fig.~\ref{fig:TransientLimit}). Already now, benefiting from the large sky coverage and realtime data transfer, ARIANNA provides useful information on transient events (see e.g. \cite{2019ATel12475}). A future detector based on ARIANNA technologies \cite{COSPAR2019} can deliver unparalleled sensitivity to transient sources within its field of view.

\subsection{Impact of Site Location on Analysis}
\label{sec:SiteImpact}

The deployment of the ARIANNA demonstrator station SP-1 approximately \SI{2.5}{km} from the Amundsen-Scott Station at the South Pole (see Sec.~\ref{sec:SouthPole} for more detail) creates the opportunity to assess the effect of the different RF backgrounds on analysis efficiency.  For this purpose, data from runs where SP-1 (South Pole) and the cosmic-ray station at Site X (Moore's Bay) were triggering on their downward facing antennas (i.e., searching for upward propagating radio signals from neutrinos) was selected. The data of both stations were analyzed by the same methods discussed above. These two stations were chosen for direct comparison because they are both constructed using an updated 8 channel version of the ARIANNA DAq board.

\begin{figure}
  \begin{centering}
        \includegraphics[width=0.49\textwidth]{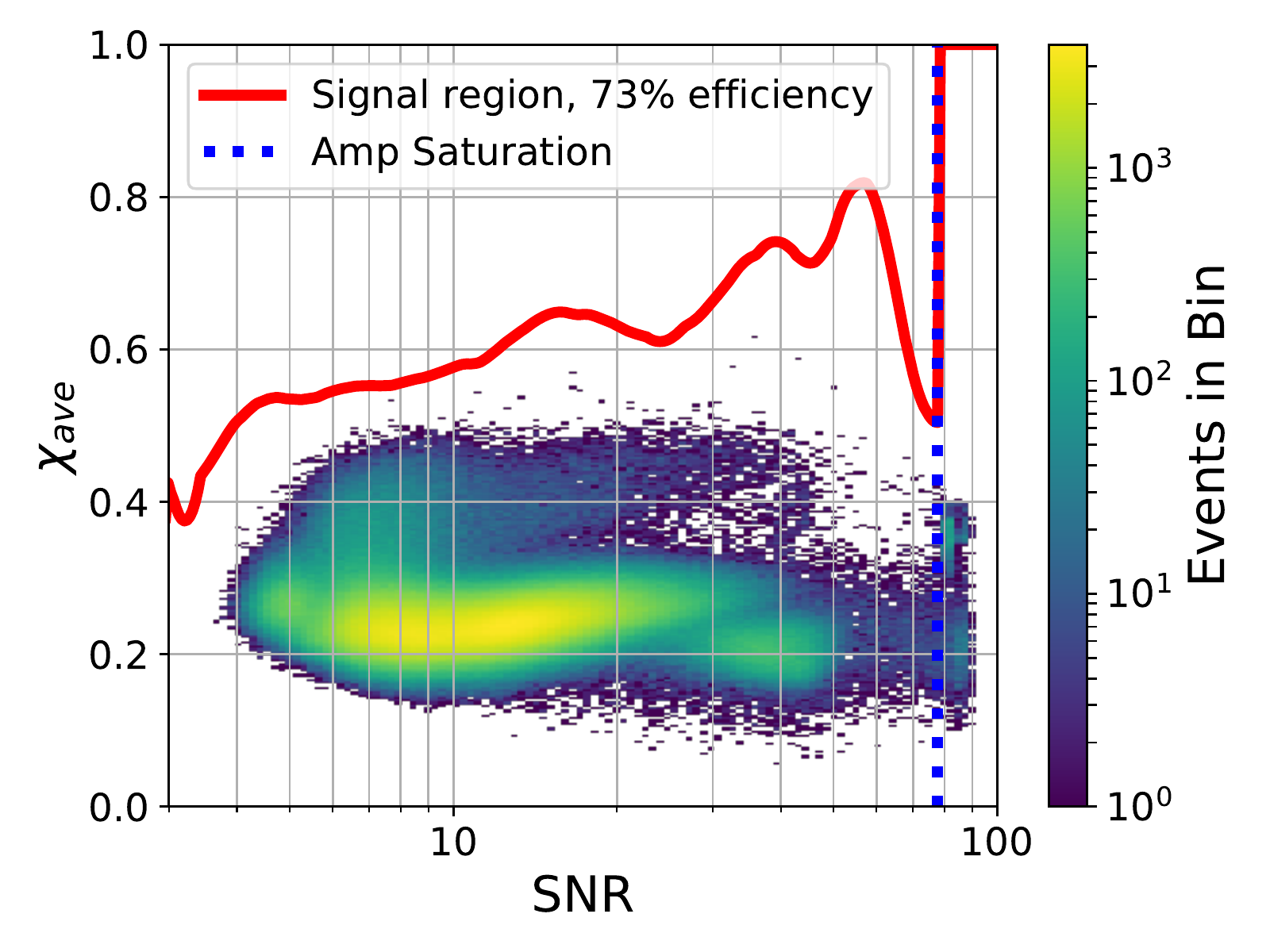}
        \includegraphics[width=0.49\textwidth]{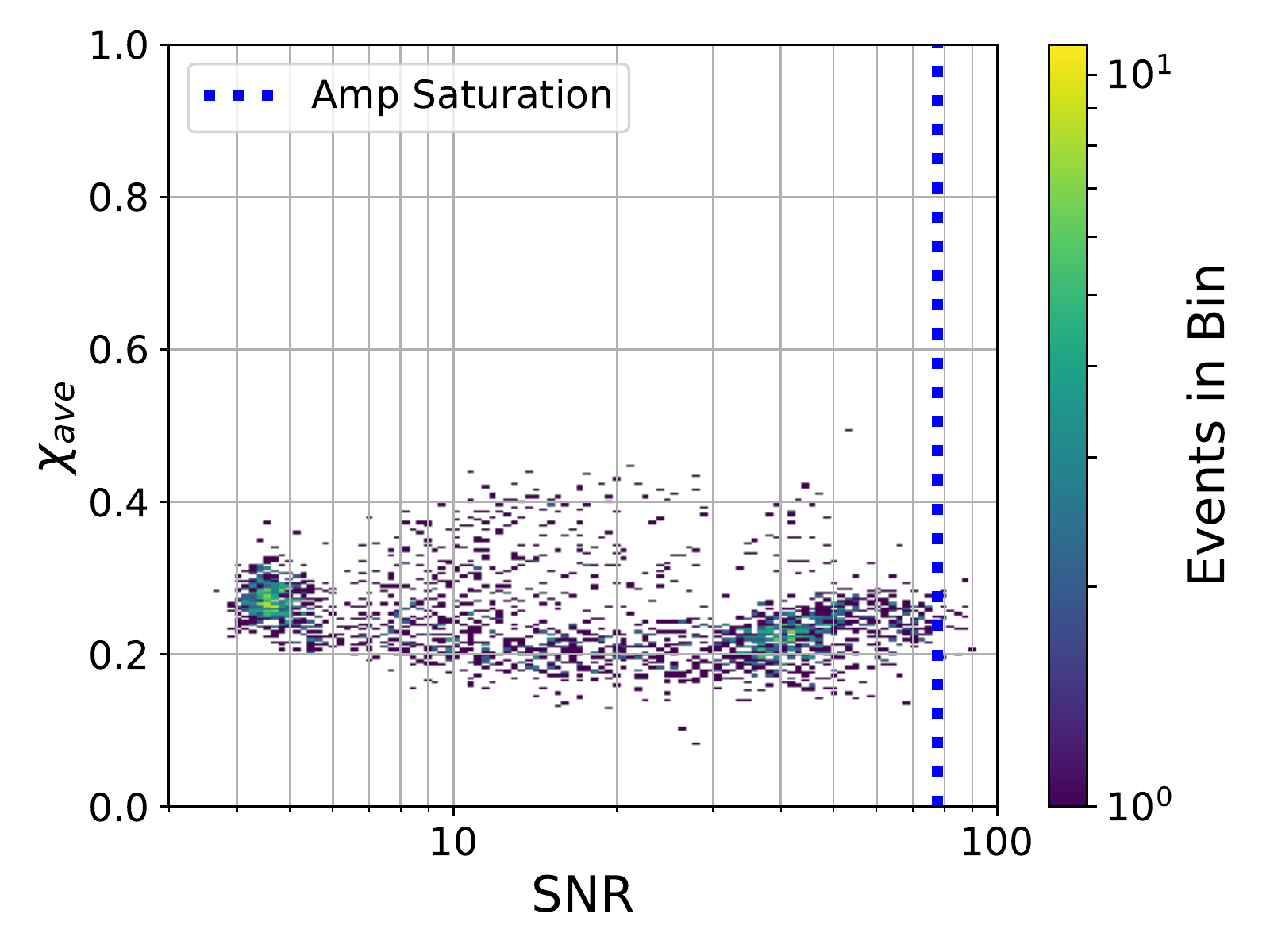}
    \caption{Distribution of triggered events for two 8 channel ARIANNA stations.  SP-1 (left) is located near the South Pole, while the station at Site X (right) is located at the Moore's bay site.  The signal region is drawn using the same procedure as shown in Sec.~\ref{sec:SignalCut}.  The relative lack of low SNR, high $\chi_{ave}$ events at Site X is evidence that the radio-quiet environment in Moore's Bay is advantageous in this analysis.}
    \label{fig:TriggeredDistAnalysis_51}
  \end{centering}
\end{figure}

We present all triggered events in the 2D signal space ($\chi_{ave}$ vs. SNR) in Fig.~\ref{fig:TriggeredDistAnalysis_51}. The signal region is calculated using the same procedure as described in Sec.~\ref{sec:SignalCut}, which we show as red curve in Fig.~\ref{fig:TriggeredDistAnalysis_51}. We apply this signal cut to a dedicated simulation data set of SP-1 and find that SP-1 achieves a signal efficiency of 73\% despite its close proximity to the Amundsen-Scott Station.  A decommissioned ARA wind turbine also contributes to the background radio noise during wind gusts.  These events were identified due to their consistent waveform properties and their arrival direction.

The same type of station at Site X in Moore's Bay was operated for less that 40 days in neutrino trigger mode, so there are a limited number of triggered events to examine and compare.  However, the distribution lacks any significant number of events with a value of $\chi_{ave} > 0.4$ and $SNR < 10$, so the expected analysis efficiency for a neutrino search would be significantly higher.

\section{Conclusion}
\label{sec:conclusion}

The operation of the ARIANNA test bed at Moore's Bay, since its completion in November 2014, has provided a robust proof of concept for an in-situ, radio-based UHE neutrino detector.  Iterative improvements  in data acquisition and power systems since the first ARIANNA prototypes \cite{ARIANNAPrototype} and initial test bed stations \cite{2015ARIANNATechnologyPaper} have improved reliability and livetime. We have presented the analysis of 4.5 years of data, leading to a 90\% confidence upper limit on the diffuse neutrino flux of $E^2\Phi=\SI{1.7e-6}{GeV cm^{-2}s^{-1}sr^{-1}}$ for a decade wide logarithmic bin centered at a neutrino energy of $10^{18}$ eV.

Since each ARIANNA station serves as an autonomous and self-contained detector, this platform can be deployed at sites in both the northern and southern hemisphere to attain full sky coverage.  Near-surface detectors allow for the easy installation of high-gain antennas in any polarization, providing increased information for precise event reconstruction. 

We have demonstrated that a simple template matching procedure is capable of rejecting all observed backgrounds over the operation of the test bed, while maintaining a combined signal efficiency of 79\% at a $4\sigma$ trigger threshold in amplitude.  Future detectors will make use of more advanced triggering and analysis techniques to push the sensitivity threshold lower, and increase the effective volume per station.  

The near-surface design of ARIANNA station provides the essential tools to contribute to the multi-messenger revolution in high energy astrophysics: reliable operation, pointing, wide-field of view, energy and real-time identification of neutrino candidates.  ARIANNA can fully participate in campaigns involving multi-wavelength (radio to gamma-ray), gravitational wave, and cosmic ray observations, with the goal of establishing both the sources of the most energetic particles known, and the mechanisms that are responsible for producing particles of extreme energy.

\section{Acknowledgements}
We are grateful to the U.S. National Science Foundation-Office of Polar Programs, the U.S. National Science Foundation-Physics Division (grant NSF-1607719) for granting the ARIANNA array at Moore's Bay. Without the invaluable support of the people at McMurdo, the ARIANNA stations would have never been built.

We acknowledge funding from the German research foundation (DFG) under grants GL 914/1-1 (CG), NE 2031/1-1, and NE 2031/2-1 (DGF, ANe, IP, CW) and the Taiwan Ministry of Science and Technology (JN, SHW).  HB acknowledges support from the Swedish Government strategic program Stand Up for Energy.  EU acknowledges support from the Uppsala university Vice-Chancellor's  travel  grant  (sponsored  by  the  Knut  and  Alice  Wallenberg  Foundation)  and  the  C.F. Liljewalch travel scholarships.  DB and ANo acknowledge support from the MEPhI Academic Excellence Project (Contract No.  02.a03.21.0005) and the Megagrant 2013 program of Russia, via agreement 14.12.31.0006 from 24.06.2013.

\bibliographystyle{JHEP}
\bibliography{HRA}

\end{document}